\documentclass[acmsmall,11pt,nonacm]{acmart}

\usepackage{amsmath,mathtools}
\usepackage{booktabs}
\usepackage{placeins}
\usepackage{float}
\usepackage{array}
\usepackage{xspace}
\usepackage{alltt}
\newenvironment{codebox}{\begin{minipage}[t]{\linewidth}\footnotesize\begin{alltt}}{\end{alltt}\end{minipage}}

\colorlet{cP}{blue!75!black}\colorlet{cC}{red!70!black}\colorlet{cB}{green!50!black}\colorlet{cZ}{violet!85!black}
\newlength{\codeH}
\newcommand{\sidelabel}[1]{\begin{minipage}[c][\codeH][c]{\linewidth}\centering\rotatebox{90}{\scriptsize\textbf{#1}}\end{minipage}}
\usepackage{tikz}
\usepackage{pgfplots}\usepgfplotslibrary{groupplots}\pgfplotsset{compat=1.17}
\usetikzlibrary{arrows.meta,positioning,calc,fit,shapes.geometric,backgrounds,decorations.pathreplacing}

\newcommand{\VH}{\textsc{VHamilton}\xspace}
\newcommand{\Ham}{\textsc{Hamilton}\xspace}
\newcommand{\FV}{\operatorname{FV}\xspace}
\newcommand{\id}{\mathrm{I}\xspace}
\newcommand{\hole}[1]{\boxed{#1}}
\newcommand{\QAnalytic}{128}
\newcommand{\coderesults}{https://github.com/JavariaGhafoor/vhamilton/tree/main/results/}
\newcommand{\codecite}[2]{\cite[\href{\coderesults #1}{\texttt{#2}}]{vhamiltoncode}}
\newcommand{\codeagda}{\cite[\href{https://github.com/JavariaGhafoor/vhamilton/tree/main/agda}{\texttt{agda}}]{vhamiltoncode}}
\newcommand{\dblcolon}{\mathbin{::}}

\title{Variational Programs with Open Clocks}
\author{Javaria Ghafoor}
\email{jghafoor@iu.edu}

\author{Amr Sabry}
\email{sabry@iu.edu}

\affiliation{%
  \department{Department of Computer Science}
  \institution{Indiana University Bloomington}
  \city{Bloomington}
  \state{Indiana}
  \country{USA}
}

\tikzset{
  wire/.style={thick},
  closedblk/.style={draw, thick, fill=gray!8, minimum height=1.25cm, inner sep=3pt, font=\footnotesize},
  openblk/.style={draw, thick, dashed, fill=violet!8, minimum height=1.25cm, inner sep=3pt, font=\footnotesize},
  smallopen/.style={draw, thick, dashed, fill=violet!8, minimum height=0.55cm, inner sep=2pt, font=\footnotesize},
  flow/.style={-{Stealth[length=5pt]}, thick},
}

\begin{document}
\begin{abstract}
Variational quantum algorithms leave the durations of Hamiltonian evolutions open for a classical optimizer to set. We present \VH, a minimal extension of the \Ham quantum programming language in which such durations are \emph{open clocks}: one program denotes a family of Hamiltonian trajectories, and a \emph{valuation} fills its clocks with exact dyadic times, so that each instance is an ordinary \Ham program that compiles to circuits as before. Because the semantics stays exact and affine in the clocks, the Hamiltonian structure of a variational program remains available for reasoning about every valuation at once. Equivalence becomes a finite calculation, which we mechanize in Agda for a fragment and use to prove the \VH form of the compiler rewrite \texttt{rzz(t)} $=$ \texttt{cx; rz(t); cx} for every clock value. We then show what the same structure supports in a two-qubit chemistry eigensolver written in \VH and run through a Haskell prototype. It supports a fusion of commuting blocks, a closed form and an expressibility proof for the ansatz, and an exact split of the error between the precision of the target and that of the clocks, with a priori bounds. It also supports a coordinate search whose queries per sweep, fixed by the spectra of the generators, do not grow with the precision. On the benchmark this search reaches the numerically verified optimum of a grid of 134 million valuations in 53 energy queries. It finds the grid optimum in only 21 of 30 validation settings, however, so the gain is conditional, not a global-optimization guarantee.
\end{abstract}
\maketitle

\section{Introduction}\label{sec:intro}
A variational quantum algorithm, such as the Variational Quantum Eigensolver (VQE) \cite{peruzzo2014} or the Quantum Approximate Optimization Algorithm (QAOA) \cite{farhi2014}, is a quantum program with holes: a classical optimizer fills the holes with numbers, runs the program, measures the energy of the state it prepares, and fills them again, so one program stands for a whole family of computations. Such algorithms are among the most promising uses of near-term quantum computers, and they raise a question familiar to programming-language researchers. Before the optimizer picks any numbers, a compiler rewrites the program to fit the hardware, so each rewrite must be correct however the holes are later filled. In Qiskit \cite{qiskit2024}, for example, matrix-based \texttt{Operator} comparison requires the holes to be filled before matrices can be compared. This motivates a representation in which equivalence can be established symbolically, before any values are chosen.

Our approach starts from the physics. A rotation gate is a snapshot of a continuous motion: the qubit turns about an axis at a steady rate, and the gate's angle records how long it turns. Physicists describe such a motion by its \emph{Hamiltonian}, which, like a velocity, says which way and how fast the state moves. The holes of a variational program are thus durations, which we call \emph{clocks}, and the optimizer sets how long each motion runs.

Seen this way, programs become easier to compare. Motions that run at the same time combine by adding their Hamiltonians, as velocities add, so a program's Hamiltonian is a sum of simple terms, each weighted by its clock: the gate it performs depends on the clocks through sines and cosines, but its Hamiltonian depends on them linearly, up to constants. Taking clocks to be exact binary fractions gives them finite representations compatible with finite-precision hardware. Two programs with holes can then be proved to denote the same Hamiltonian for every filling at once, by comparing finitely many linear expressions with exact coefficients.

\VH, the language we introduce, makes this precise. It builds on \Ham \cite{hamilton}, a recent language whose programs describe simultaneous motions with exact binary-fraction clocks (dyadic rationals), compile to circuits, and have a sound and complete equational theory: its equations prove two programs equal exactly when they are. \Ham fixes every clock in the program; \VH adds a single construct, the open clock, so one program describes a family of motions, each an ordinary \Ham program.

\begin{figure}[t]
\centering
\small
\begin{tabular}{@{}m{0.025\linewidth}@{\hspace{0.01\linewidth}}m{0.29\linewidth}@{\hspace{0.02\linewidth}}m{0.025\linewidth}@{\hspace{0.01\linewidth}}m{0.62\linewidth}@{}}
\toprule
\sidelabel{\VH} &
\begin{minipage}[c][\codeH][t]{\linewidth}
\begin{codebox}
\textrm{(a) one QAOA layer}
\textcolor{cP}{\(P\;=\;YI\sb{3}\parallel IY\sb{3}\)}
\textcolor{cC}{\(C(\gamma)=ZZ(\gamma)\)}
\textcolor{cB}{\(B(\beta)=XI(\beta)\parallel IX(\beta)\)}
ansatz = [\(P\), \(C(\gamma)\), \(B(\beta)\)]
\end{codebox}\par\vspace{3pt}\hrule\vspace{3pt}
\begin{codebox}
\textrm{(b) two clocks, noncommuting}
\textrm{terms}  \textcolor{cZ}{\(Z(\alpha)\parallel X(\beta)\)}
\end{codebox}\par\vspace{3pt}\hrule\vspace{3pt}
\begin{codebox}
\textrm{(c) a rewrite, proved for all \(\nu\)}
\textcolor{cC}{\(ZZ(\gamma)\simeq CX\,(\id\otimes Z(\gamma))\,CX\sp\dagger\)}
\end{codebox}
\end{minipage}
& \sidelabel{Qiskit} &
\begin{minipage}[c][\codeH][t]{\linewidth}
\begin{codebox}
a, b, g = Parameter("\(\alpha\)"), Parameter("\(\beta\)"), Parameter("\(\gamma\)")
qc = QuantumCircuit(2)
\textcolor{cP}{qc.ry(pi/2, [0, 1])                 # preparation}
\textcolor{cC}{qc.rzz(4*pi*g, 0, 1)                # cost}
\textcolor{cB}{qc.rx(4*pi*b, 0); qc.rx(4*pi*b, 1)  # mixer}
\end{codebox}\par\vspace{3pt}\hrule\vspace{3pt}
\begin{codebox}
qc = QuantumCircuit(1); N = 2**n
\textcolor{cZ}{for _ in range(N): qc.rz(4*pi*a/N, 0); qc.rx(4*pi*b/N, 0)}
\end{codebox}\par\vspace{3pt}\hrule\vspace{3pt}
\begin{codebox}
\textcolor{cC}{qc1.rzz(4*pi*g, 0, 1)}
\textcolor{cC}{qc2.cx(0, 1); qc2.rz(4*pi*g, 1); qc2.cx(0, 1)}
Operator(qc1) == Operator(qc2)  # raises TypeError
\end{codebox}
\end{minipage}\\
\bottomrule
\end{tabular}
\caption{\VH processes and their Qiskit realizations: (a) one QAOA layer; (b) a Trotter loop for noncommuting terms; (c) an equivalent rewrite of the cost block.}
\Description{Side-by-side code. Left: VHamilton definitions of a QAOA layer and of a two-clock block. Right: the corresponding Qiskit circuits, with explicit angle conversions and an explicit Trotter loop. Third row: the rewrite of the cost block, proved equivalent for every valuation in VHamilton, while Qiskit's Operator comparison fails on the unbound parameter.}
\label{fig:code}
\end{figure}

The right column of Fig.~\ref{fig:code} shows how such programs are written today, in Qiskit \cite{qiskit2024}; the left column writes them in \VH, with corresponding lines in the same colour. Part (a) is one QAOA layer for MaxCut on a single edge: a preparation $P$, a cost $C(\gamma)$ and a mixer $B(\beta)$. \VH writes each as the motion it describes, with its clock: $YI_3$ turns the first qubit about $Y$ for the fixed time $2^{-3}$, given by its subscript; $ZZ(\gamma)$ couples the two qubits for the open time $\gamma$; and $\parallel$ runs motions together, adding their Hamiltonians.

Qiskit instead lists gates, whose angles the programmer derives from these times by
\[
R_P(\theta)=e^{-i\theta P/2},\qquad \theta=4\pi t
\]
for a motion run for time $t$, and the mixer's one field appears as two gates sharing a parameter.

Part (b) asks for two noncommuting motions at once: \VH denotes their combined motion and leaves Trotterization to the compiler \cite{trotter1959}, while Qiskit hard-codes it as a loop of $2^n$ pairs of short turns.

The payoff appears in Fig.~\ref{fig:code}c. On hardware that cannot couple two qubits through $ZZ$ directly, a compiler rewrites the cost $C(\gamma)$ as a $Z$ rotation between two controlled-NOT ($CX$) gates. Qiskit's \texttt{Operator} comparison refuses to check this rewrite while \texttt{g} is unbound, but in \VH both forms denote the same Hamiltonian, $\gamma\,Z\otimes Z$, so comparing their finitely many coefficients proves the rewrite for every value of $\gamma$, a check that Agda carries out by computation (Sec.~\ref{sec:equivalence}).

The advantage of leaving the Hamiltonian visible extends beyond equivalence checking. In our chemistry eigensolver, the same representation exposes commuting blocks that can be fused, separates the errors introduced by the precision of the target and that of the clocks, and reveals the generator spectra that fix the shape of each clock's energy curve. Open clocks thus support not only verifying a variational program but also reasoning about how to compile and optimize it.

Secs.~\ref{sec:hamilton}--\ref{sec:examples} recall \Ham, add open clocks and write variational programs; Sec.~\ref{sec:equivalence} proves the rewrite in Agda \cite{agda}, Sec.~\ref{sec:running} runs the programs, Sec.~\ref{sec:chem} writes a chemistry eigensolver in \VH and accounts for its errors, and Sec.~\ref{sec:scaling} shows how the structure of a program can replace exhaustive search as clocks multiply.

\section{Fixed Clocks in \Ham}\label{sec:hamilton}
Quantum programs are usually written as circuits: an itinerary of gates, each a unitary applied after the one before. A Hamiltonian $H$, by contrast, drives a state continuously along the trajectory
\[
t\mapsto e^{-2\pi itH},
\]
and a gate is where such a trajectory is at one moment. \Ham \cite{hamilton} makes the trajectories themselves programs: parallel interactions add their generators, $H_1\parallel H_2\mapsto H_1+H_2$, and ordered gates arise only when the motion is compiled.

We recall \Ham here, where every clock is fixed, before leaving clocks open in \VH. It has four kinds of terms, with $k\in\mathbb N\cup\{0\}$:
\[
\begin{array}{@{}l@{\qquad}r@{\;}c@{\;}l@{}}
\text{Transformations} & u & ::= & \id \mid H \mid S \mid CX \mid SWAP \mid u\cdot u \mid u\otimes u\\
\text{Bare actions} & d & ::= & 0 \mid \delta_k \mid -\delta_k \mid d\oplus d\\
\text{Actions} & a & ::= & d \mid uau^\dagger\\
\text{Processes} & p & ::= & a \mid p\parallel p
\end{array}
\]

The scalars $\delta_k$, $-\delta_k$ and $0$ move forward, backward and not at all; $\oplus$ arranges them into a diagonal motion, one per basis state.

Transformations $u$ are Clifford operations, and processes $p$ run actions in parallel. Types are positive integers, the Hilbert-space dimensions: $\id,H,S:2$, $CX,SWAP:4$ and $0,\pm\delta_k:1$; $\otimes$ multiplies dimensions, $\oplus$ adds them, and composition, conjugation and $\parallel$ require matching dimensions. The meaning of a process is its Hamiltonian:
\[
\begin{array}{@{}l@{\qquad}r@{\;}c@{\;}l@{}}
\text{Scalars are dyadic numbers} & \llbracket\delta_k\rrbracket_H & = & 2^{-k}\\
\text{Direct sums stay direct sums} & \llbracket d_1\oplus d_2\rrbracket_H & = & \llbracket d_1\rrbracket_H\oplus\llbracket d_2\rrbracket_H\\
\text{Conjugation changes basis} & \llbracket uau^\dagger\rrbracket_H & = & \llbracket u\rrbracket_U\llbracket a\rrbracket_H\llbracket u\rrbracket_U^\dagger\\
\text{Parallel composition adds} & \llbracket p\parallel q\rrbracket_H & = & \llbracket p\rrbracket_H+\llbracket q\rrbracket_H
\end{array}
\]

A process runs for one unit of time, and its unitary is where it ends:
\[
\gamma_p(t)=e^{-2\pi it\llbracket p\rrbracket_H},\qquad \llbracket p\rrbracket_U=\gamma_p(1).
\] \noindent Running $H$ for time $t$ equals running $tH$ for one unit \cite{hamilton}, so coefficients are effective durations, equivalently normalized rotation angles. We call each scalar $\pm\delta_k$ a \emph{clock}: it sets how long, in effective time, the motion it scales runs. The examples below show what such clocks express in \Ham.

\begin{figure}[t]
\centering
\begin{tikzpicture}[font=\scriptsize, >={Stealth[length=3pt]}]
  \foreach \i/\k/\gate/\lab in {0/0/{\id}/{4\pi},1/1/{-\id}/{2\pi},2/2/{Z}/{\pi},3/3/{S}/{\pi/2},4/4/{T}/{\pi/4}} {
    \begin{scope}[shift={(\i*0.72,0)}]
      \draw[gray!45] (0,0) circle (0.29);
      \fill (0,0) circle (0.8pt);
      \node at (0,0.52) {$Z_{\k}$};
      \node at (0,-0.48) {$\lab$};
      \node at (0,-0.74) {$\gate$};
    \end{scope}
  }
  \draw[->, thick] plot[domain=0:712, samples=140, smooth] ({0.17*(1+0.6*\x/720)*sin(\x)}, {0.17*(1+0.6*\x/720)*cos(\x)});
  \draw[->, thick] (0.72,0.24) arc[start angle=90, end angle=-262, radius=0.24];
  \draw[->, thick] (1.44,0.24) arc[start angle=90, end angle=-90, radius=0.24];
  \draw[->, thick] (2.16,0.24) arc[start angle=90, end angle=0, radius=0.24];
  \draw[->, thick] (2.88,0.24) arc[start angle=90, end angle=45, radius=0.24];
  \node[font=\footnotesize\bfseries] at (-0.5,0.52) {(a)};
  \begin{scope}[shift={(4.05,0)}]
    \draw[->, thick] (0,-0.6) -- (0,0.6) node[above, inner sep=1pt]{$Z$};
    \draw[thick, ->] (-0.3,0.2) arc[start angle=180, end angle=335, x radius=0.3, y radius=0.1];
    \node at (0,-0.78) {$Z_2$};
    \draw[->, gray] (0.4,0) -- node[above]{$H{\cdot}H^\dagger$} (0.95,0);
    \draw[->, thick] (1.0,0) -- (1.85,0) node[right, inner sep=1pt]{$X$};
    \draw[thick, ->] (1.625,0.3) arc[start angle=90, end angle=245, x radius=0.1, y radius=0.3];
    \node at (1.625,-0.78) {$X_2$};
    \node[font=\footnotesize\bfseries] at (-0.3,0.52) {(b)};
  \end{scope}
  \begin{scope}[shift={(7.10,0)}]
    \foreach \i/\ket/\entry/\dir in {0/{00}/{\delta_3}/-1, 1/{01}/{-\delta_3}/1, 2/{10}/{-\delta_3}/1, 3/{11}/{\delta_3}/-1} {
      \begin{scope}[shift={(\i*0.66,0)}]
        \draw[gray!45] (0,0) circle (0.27);
        \fill (0,0) circle (0.8pt);
        \draw[->, thick] (0,0.22) arc[start angle=90, delta angle={\dir*45}, radius=0.22];
        \node at (0,0.5) {$|\ket\rangle$};
        \node at (0,-0.48) {$\entry$};
      \end{scope}
    }
    \node at (2.42,0) {$\propto$};
    \begin{scope}[shift={(2.62,0)}]
      \draw[thick] (0,0.25) -- (1.25,0.25);
      \draw[thick] (0,-0.25) -- (1.25,-0.25);
      \draw[thick] (0.3,0.25) -- (0.3,-0.25);
      \fill (0.3,0.25) circle (1.8pt);
      \fill (0.3,-0.25) circle (1.8pt);
      \foreach \y in {0.25,-0.25}
        \node[draw, thick, fill=white, minimum size=0.34cm, inner sep=0pt] at (0.85,\y) {$S$};
    \end{scope}
    \node[font=\footnotesize\bfseries] at (-0.5,0.52) {(c)};
  \end{scope}
  \begin{scope}[shift={(11.65,-0.55)}, x=1.1cm, y=1.1cm]
    \draw[dashed, thick] (0,0) -- (1,1);
    \draw[black!25, line width=1.1pt] (0,0) -- (1,0) -- (1,1);
    \draw[black!55, line width=1.1pt] (0,0) -- (0.5,0) -- (0.5,0.5) -- (1,0.5) -- (1,1);
    \draw[black, thick] (0,0) \foreach \i in {1,...,4} { -- ({\i/4},{(\i-1)/4}) -- ({\i/4},{\i/4}) };
    \node[below, inner sep=1.5pt] at (0.5,0) {$X$};
    \node[right, inner sep=1.5pt] at (1,0.5) {$Z$};
    \node[font=\footnotesize\bfseries] at (-0.3,0.97) {(d)};
  \end{scope}
\end{tikzpicture}
\caption{Fixed clocks: (a) $Z_k$, (b) conjugation, (c) the entangler $ZZ_3$, (d) Trotter levels of $X_3\parallel Z_3$.}
\Description{Clock dials for Z_0 to Z_4; an axis turned from Z to X by Hadamard conjugation; four phase dials for the two-qubit basis states with the equivalent circuit CZ followed by S on each qubit; Trotter staircases approaching a diagonal.}
\label{fig:fixed}
\end{figure}
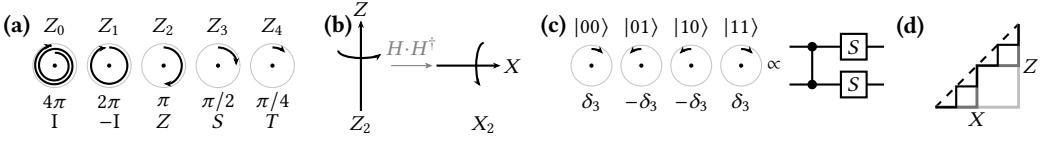

\textbf{Turning about one axis:} Let
\[
Z_k:=\delta_k\oplus(-\delta_k),\qquad\text{with Hamiltonian}\qquad 2^{-k}Z.
\]
It rotates about the $Z$-axis by $4\pi/2^k$; each step in $k$ halves the turn (Fig.~\ref{fig:fixed}a). $Z_0,Z_1$ make two and one full turns, while $Z_2,Z_3,Z_4$ give $Z,S,T$, up to phase. Conjugation changes the axis: $X_2:=H\,Z_2\,H^\dagger$ makes the same half turn about $X$, and its unit-time evolution is the $X$ gate, up to phase (Fig.~\ref{fig:fixed}b).

\textbf{Entangling two qubits:} The action
\[
ZZ_3:=\delta_3\oplus(-\delta_3)\oplus(-\delta_3)\oplus\delta_3\qquad\text{has Hamiltonian}\qquad\tfrac18\,Z\otimes Z.
\]
Its clocks run forward on $|00\rangle,|11\rangle$ and backward on $|01\rangle,|10\rangle$; it ends at $CZ$ followed by $S\otimes S$, up to phase (Fig.~\ref{fig:fixed}c).

\textbf{Two axes at once:} The process
\[
X_3\parallel Z_3\qquad\text{has Hamiltonian}\qquad\tfrac18\,(X+Z),
\]
rotating about the axis halfway between $X$ and $Z$. The compiler uses Lie--Trotter interleaving \cite{trotter1959}: level $n$ alternates $2^n$ short turns about each axis, approaching the exact motion that $X_3\parallel Z_3$ denotes; in Fig.~\ref{fig:fixed}d, levels $n=0,1,2$ (light to dark) approach that motion (dashed).

In every example the clocks are written into the program: $S$ and $T$ are \emph{different programs}, so how far to travel is fixed when the program is written, which is exactly the decision variational algorithms must postpone.

\section{Open Clocks and Valuations}\label{sec:valuations}
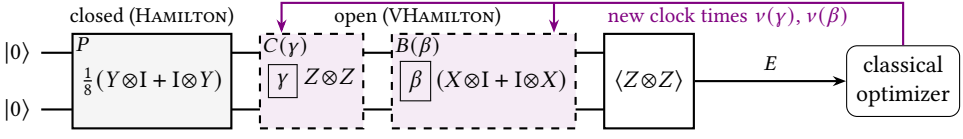
\begin{figure}[t]
\centering
\begin{tikzpicture}[x=1cm,y=1cm]
  \draw[wire] (0,0) -- (8.6,0);
  \draw[wire] (0,-0.75) -- (8.6,-0.75);
  \node[left, font=\footnotesize] at (0,0) {$|0\rangle$};
  \node[left, font=\footnotesize] at (0,-0.75) {$|0\rangle$};
  \node[closedblk, anchor=west] (P) at (0.4,-0.375) {$\tfrac18(Y\!\otimes\!\id+\id\!\otimes\! Y)$};
  \node[openblk, anchor=west] (C) at ($(P.east)+(0.35,0)$) {$\hole{\gamma}\;Z\!\otimes\! Z$};
  \node[openblk, anchor=west] (B) at ($(C.east)+(0.35,0)$) {$\hole{\beta}\,(X\!\otimes\!\id+\id\!\otimes\! X)$};
  \node[draw, thick, fill=white, minimum height=1.25cm, font=\footnotesize, anchor=west] (M) at ($(B.east)+(0.35,0)$) {$\langle Z\!\otimes\! Z\rangle$};
  \node[draw, rounded corners, font=\footnotesize, align=center, anchor=west] (O) at ($(M.east)+(2.0,0)$) {classical\\optimizer};
  \draw[flow] (M) -- node[above, font=\scriptsize]{$E$} (O);
  \node[anchor=north west, font=\scriptsize, inner sep=1.5pt] at (P.north west) {$P$};
  \node[anchor=north west, font=\scriptsize, inner sep=1.5pt] at (C.north west) {$C(\gamma)$};
  \node[anchor=north west, font=\scriptsize, inner sep=1.5pt] at (B.north west) {$B(\beta)$};
  \coordinate (top) at (0,0.66);
  \coordinate (Cin) at ($(C.north west)+(0.3,0)$);
  \coordinate (Bin) at ($(B.north east)+(-0.3,0)$);
  \draw[flow, violet] (O.north) -- (O.north |- top) -- (Cin |- top) -- (Cin);
  \draw[flow, violet] (Bin |- top) -- (Bin);
  \node[font=\scriptsize, anchor=north, inner sep=1.5pt] at (P.north |- top) {closed (\Ham)};
  \node[font=\scriptsize, anchor=north, inner sep=1.5pt] at ($(Cin |- top)!0.5!(Bin |- top)$) {open (\VH)};
  \node[font=\scriptsize, anchor=north, inner sep=1.5pt, violet] at ($(Bin |- top)!0.5!(O.north |- top)$) {new clock times $\nu(\gamma),\nu(\beta)$};
\end{tikzpicture}
\caption{A variational ansatz as a circuit with holes: one QAOA layer for MaxCut on a single edge.}
\Description{A two-qubit circuit: a closed preparation block, two dashed open blocks with boxed clocks gamma and beta, a measurement of ZZ, and an optimizer feeding new clock times back into the holes.}
\label{fig:holes}
\end{figure}

A variational algorithm fixes the shape of its circuit and leaves the durations to the optimizer: QAOA, for instance, alternates evolution under a cost Hamiltonian for time $\gamma$ and under a mixer for time $\beta$. Fig.~\ref{fig:holes} draws the QAOA layer of Fig.~\ref{fig:code}a as a circuit with holes. Each block runs its Hamiltonian for unit time: the closed block has fixed clocks and prepares $|{+}{+}\rangle$, while the open blocks leave their clocks $\gamma,\beta$ to the optimizer. In \Ham a duration is a fixed $\delta_k$; \VH turns it into a hole. The minimal extension adds open clocks to the bare actions: for a hole named $\alpha\in\mathcal V$, a set of variables,
\[
d ::= \dots \mid \delta_k^{\alpha} \mid -\delta_k^{\alpha},\qquad \delta_k^{\alpha},\,-\delta_k^{\alpha}:1.
\]

Nothing else changes in the syntax. We call $\pm\delta_k^{\alpha}$ an \emph{open clock}, named by its variable $\alpha$: all occurrences of $\alpha$ refer to the same hole in the circuit picture.

What is new is the \emph{valuation}. A valuation $\nu:\FV(p)\to\mathbb N\cup\{0\}$ fills the holes of $p$, where \(\FV(p)\subseteq\mathcal V\) denotes the finite set of variables occurring in \(p\). \Ham has no counterpart: there, every clock is already a number.

A filled hole behaves like a fixed clock. The semantics $\llbracket p\rrbracket_H^\nu$ is \Ham's, clause for clause, with $\nu$ passed along, plus the clause
\[
\llbracket\pm\delta_k^{\alpha}\rrbracket_{H}^{\nu}=\pm\frac{\nu(\alpha)}{2^k},
\qquad\text{so that}\qquad
\llbracket p\rrbracket_U^\nu=\gamma_{p,\nu}(1)=e^{-2\pi i\llbracket p\rrbracket_H^\nu}.
\]
A valuation sets an effective duration (zero disables an interaction); the sign in the syntax sets the direction. A \VH program thus denotes a \emph{family} of trajectories, $\nu\mapsto\gamma_{p,\nu}$: the syntax fixes which motions occur and the valuation fixes how long each runs.

Fixing $\nu$ makes every coefficient a concrete dyadic number, so each instance can be written in \Ham's syntax, as an ordinary \Ham program, and compiled as before.

\textbf{A finite family of programs.} A Qiskit \texttt{Parameter} stands for any real angle, so a parameterized circuit denotes a continuum of circuits. In \VH a valuation may give a clock any natural number, but restricting each of the $m$ clocks to the precision-$r$ grid $\nu(\alpha)\in\{0,\dots,2^r-1\}$ gives $2^{mr}$ valuations ($2^{m(r-1)}$ on the one-period QAOA grid of Sec.~\ref{sec:running}). Each instantiates an ordinary \Ham program. So the optimum at precision $r$ is well defined, and exhaustive search, which evaluates every valuation, finds it up to the floating-point error of each computed energy.

Precision nests, $q/2^r=2q/2^{r+1}$, so a new bit loses no valuation. And every member lies within \Ham's complete equational calculus \cite{hamilton}, so a statement such as ``the optimized cost block is $ZZ_3$'' (Sec.~\ref{sec:running}) is an equation between programs.

\section{What Open Clocks Express}\label{sec:examples}
The examples below show what open clocks let a variational program say.

\begin{figure}[t]
\centering
\begin{tikzpicture}[font=\scriptsize, >={Stealth[length=3pt]}]
  \begin{scope}
    \draw[gray!45] (0,0) circle (0.42);
    \fill (0,0) circle (0.8pt);
    \draw[->, thick, dashed, violet] (0,0.32) arc[start angle=90, end angle=-235, radius=0.32];
    \foreach \a/\g/\n in {45/T/1, 0/S/2, -90/Z/4, 90/{-\id}/8} {
      \draw[thick] (\a:0.38) -- (\a:0.47);
      \node at (\a:0.68) {$\g$};
    }
    \node at (0,-0.98) {$Z(\alpha)$, $\nu(\alpha)=1,2,4,8$};
    \node[font=\footnotesize\bfseries] at (-0.85,0.62) {(a)};
  \end{scope}
  \begin{scope}[shift={(2.55,0)}]
    \node[draw, thick, dashed, fill=violet!8, inner sep=2pt] (h) at (0,0) {$\beta$};
    \foreach \y/\lab in {0.32/{XI}, -0.32/{IX}} {
      \draw[thick] (0.75,\y) -- (2.05,\y) node[right, inner sep=1pt]{$\lab$};
      \draw[thick, ->] (1.4,\y+0.17) arc[start angle=90, end angle=410, x radius=0.07, y radius=0.17];
      \draw[violet, thick] (h.east) -- (1.3,\y);
    }
    \node at (1.0,-0.98) {$B(\beta)$: one name, two fields};
    \node[font=\footnotesize\bfseries] at (-0.45,0.62) {(b)};
  \end{scope}
  \begin{scope}[shift={(5.95,0)}]
    \foreach \i/\entry/\dir/\open in {0/{\delta_k^\alpha}/-1/1, 1/{-\delta_k^\alpha}/1/1, 2/{0}/0/0, 3/{0}/0/0} {
      \begin{scope}[shift={(\i*0.72,0)}]
        \draw[gray!45] (0,0) circle (0.27);
        \fill (0,0) circle (0.8pt);
        \ifnum\open=1
          \draw[->, thick, dashed, violet] (0,0.21) arc[start angle=90, delta angle={\dir*100}, radius=0.21];
        \fi
        \node at (0,-0.52) {$\entry$};
      \end{scope}
    }
    \node at (0.36,0.5) {branch $|0\rangle$};
    \node at (1.80,0.5) {branch $|1\rangle$};
    \draw[->, gray] (2.45,0) -- node[below, inner sep=2pt]{$SH{\cdot}(SH)^\dagger$} (3.75,0);
    \begin{scope}[shift={(3.9,0)}]
      \draw[thick] (0,0.25) -- (1.3,0.25);
      \draw[thick] (0,-0.25) -- (1.3,-0.25);
      \draw[thick] (0.8,0.19) -- (0.8,-0.09);
      \draw[thick, fill=white] (0.8,0.25) circle (2pt);
      \node[draw, thick, dashed, fill=violet!8, inner sep=1.5pt, minimum height=0.32cm] at (0.8,-0.25) {$\alpha\,Y$};
    \end{scope}
    \node at (2.45,-0.98) {$CY(\alpha)$: zero entries freeze branch $|1\rangle$};
    \node[font=\footnotesize\bfseries] at (-0.5,0.62) {(c)};
  \end{scope}
\end{tikzpicture}
\caption{Open clocks: (a) one program, many gates; (b) one clock, two fields; (c) a controlled clock.}
\Description{A single dial with an open dashed hand and ticks for T, S, Z and -I; one boxed clock beta wired to rotations on two qubits; four phase dials of which only the first branch carries open clocks, next to a controlled-Y circuit with an open clock.}
\label{fig:open}
\end{figure}
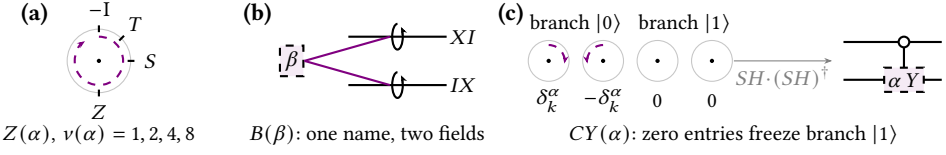

\textbf{One program, a ladder of gates:}
\[
Z(\alpha):=\delta_k^\alpha\oplus(-\delta_k^\alpha)\qquad\text{has Hamiltonian}\qquad\frac{\nu(\alpha)}{2^k}\,Z.
\]
At $k=4$ it is the rotation $R_z(\pi\nu(\alpha)/4)$, and the valuations $\nu(\alpha)=1,2,4,8$ give $T$, $S$, $Z$ and $\id$, up to phase. The four \Ham programs $Z_4,Z_3,Z_2,Z_1$ of Sec.~\ref{sec:hamilton} become four settings of one program, and every value in between is a further rotation (Fig.~\ref{fig:open}a).

\textbf{One clock, many interactions:} A variable may occur more than once. The mixer
\[
B(\beta):=XI(\beta)\parallel IX(\beta),\qquad\text{with Hamiltonian}\qquad\frac{\nu(\beta)}{2^k}\,(X\otimes\id+\id\otimes X),
\]
drives a field on each of two qubits from one clock $\beta$ (Fig.~\ref{fig:open}b). Here $XI(\beta)$ is the $H\otimes\id$ conjugate of $\delta_k^\beta\oplus\delta_k^\beta\oplus(-\delta_k^\beta)\oplus(-\delta_k^\beta)$; $IX(\beta)$ is analogous. Sharing needs no new construct, and because the two terms commute, the block compiles exactly.

Both Qiskit and \VH share one parameter across interactions, but \VH makes the resulting simultaneous Hamiltonian field explicit. Contrast $Z(\alpha)\parallel X(\beta)$ of Fig.~\ref{fig:code}b: two clocks on noncommuting terms, whose ratio sets the rotation axis and whose evolution our compiler approximates by Trotterization.

\textbf{A clock behind a control:} Zero entries freeze part of the space. The action
\[
CY(\alpha):=(\id\otimes SH)\big(\delta_k^\alpha\oplus(-\delta_k^\alpha)\oplus 0\oplus 0\big)(\id\otimes SH)^\dagger
\]
has Hamiltonian
\[
\frac{\nu(\alpha)}{2^k}\,|0\rangle\langle0|\otimes Y.
\]
It turns the second qubit only when the first is $|0\rangle$. That is a controlled rotation, with no control construct (Fig.~\ref{fig:open}c).

\textbf{The QAOA layer:} Returning to Fig.~\ref{fig:code}a, the left column combines these constructions into a variational ansatz: $YI_3$ and $IY_3$ denote $\tfrac1{2^3}Y\otimes\id$ and $\tfrac1{2^3}\id\otimes Y$, $C(\gamma)=ZZ(\gamma)$ is $ZZ_3$ with each $\pm\delta_3$ replaced by $\pm\delta_k^\gamma$, where $k$ sets the clock's resolution, and $B(\beta)$ is the shared-clock mixer defined earlier. The ansatz runs these blocks in sequence, preparing
\[
|\psi(\nu)\rangle=\llbracket B(\beta)\rrbracket_U^\nu\,\llbracket C(\gamma)\rrbracket_U^\nu\,\llbracket P\rrbracket_U\,|00\rangle,
\]
where $P$ takes $|00\rangle$ to $|{+}{+}\rangle$. \VH itself has no sequencing: each block is one process, the ansatz fixes their order outside the language, and $\parallel$ only ever adds generators, within a block. The noncommuting example of Fig.~\ref{fig:code}b is also one process,
\[
Z(\alpha)\parallel X(\beta),\qquad\text{with Hamiltonian}\qquad\frac{\nu(\alpha)}{2^k}\,Z+\frac{\nu(\beta)}{2^k}\,X,
\]
where the Qiskit loop uses \texttt{a} and \texttt{b} for $\nu(\alpha)/2^k$ and $\nu(\beta)/2^k$. That loop is written by hand because Qiskit's \texttt{PauliEvolutionGate} supports a symbolic time but not independent symbolic Pauli coefficients (v2.6).

Open clocks let one program describe a family of motions; we next ask whether two such families are equivalent without fixing their clocks.

\section{Equivalence for Every Valuation}\label{sec:equivalence}
Compilers and optimizers rewrite programs, so they need to know when two variational programs agree for every setting of their parameters. This parameterized equivalence has been approached with a ZX-calculus procedure on symbolic phases combined with instantiation \cite{peham2023}, symbolic tensor decision diagrams \cite{hong2024}, rewrites verified by a theorem prover in Quartz \cite{xu2022quartz}, a Coq-verified optimizer in VOQC \cite{hietala2021}, and path sums \cite{amy2019}. These approaches establish circuit identities through graphical, matrix-based, decision-diagram, or other symbolic representations.

We instead reason directly about the additive Hamiltonians denoted by open processes, before exponentiation or circuit compilation. Recall the rewrite of Fig.~\ref{fig:code}c, which holds since conjugating $Z$ on the second qubit by $CX$ gives $Z\otimes Z$. In \VH, the same two programs are written over one clock $\gamma$:
\[
ZZ(\gamma):=\delta_k^\gamma\oplus(-\delta_k^\gamma)\oplus(-\delta_k^\gamma)\oplus\delta_k^\gamma,
\qquad
CX\,\big(\delta_k^\gamma\oplus(-\delta_k^\gamma)\oplus\delta_k^\gamma\oplus(-\delta_k^\gamma)\big)\,CX^\dagger.
\]
The second is $\id\otimes Z(\gamma)$ conjugated by $CX$.

Because a valuation only scales clocks, each Pauli coefficient of $\llbracket p\rrbracket_H^\nu$ is an affine expression in the clock values $\nu(\alpha)/2^k$, with exact dyadic coefficients. Equality of these finitely many expressions establishes equality of the Hamiltonians for every valuation, and hence of their induced evolution unitaries. We call this \emph{Hamiltonian equivalence}; it is stronger than unitary equivalence, since distinct Hamiltonians can share a unit-time exponential.

\begin{figure}[t]
\centering
\small
\newcommand{\g}{\textcolor{cC}{\gamma}}%
\begin{tabular}{@{}p{0.97\linewidth}@{}}
\toprule
\textbf{(1) Expand} each program into Pauli strings, by $d_0\oplus d_1=\id\otimes\frac{d_0+d_1}2+Z\otimes\frac{d_0-d_1}2$; so $\delta_k^\gamma\oplus-\delta_k^\gamma=\gamma Z$, with $\gamma$ a symbol for $\nu(\gamma)/2^k$, $\id Z$ for $\id\otimes Z$, and conjugation by $CX$ maps $\id Z\mapsto ZZ$:
\[
\begin{aligned}
ZZ(\gamma)&=\gamma Z\oplus(-\gamma Z)&&=\id\otimes 0+Z\otimes\gamma Z&&=\gamma\,ZZ\\
CX\,(\id\otimes Z(\gamma))\,CX^\dagger&=CX\,(\gamma Z\oplus\gamma Z)\,CX^\dagger&&=CX\,(\gamma\,\id Z)\,CX^\dagger&&=\gamma\,ZZ
\end{aligned}
\]
\textbf{(2) Compare} all 16 coefficients, \texttt{p \(\simeq\) q = \(\forall\) Q \(\to\) coef \(\llbracket\) p \(\rrbracket\) Q \(\equiv\) coef \(\llbracket\) q \(\rrbracket\) Q}:\\[2pt]
\centering\footnotesize\setlength{\tabcolsep}{3pt}%
\begin{tabular}{@{}r@{\hspace{5pt}}*{4}{w{c}{1.2em}}@{\hspace{6pt}}c@{\hspace{6pt}}*{4}{w{c}{1.2em}}@{\hspace{8pt}}l@{}}
\texttt{\(\llbracket\)ZZ\(\rrbracket\):} & $\id$ & $X$ & $Y$ & $Z$ & \texttt{\(\llbracket\)CX-IZ-CX\(\rrbracket\):} & $\id$ & $X$ & $Y$ & $Z$ & \\\cmidrule(lr){2-5}\cmidrule(lr){7-10}
$\id$ & 0&0&0&0 & & 0&0&0&0 & \texttt{refl}\\
$X$ & 0&0&0&0 & $=$ & 0&0&0&0 & \texttt{refl}\\
$Y$ & 0&0&0&0 & & 0&0&0&0 & \texttt{refl}\\
$Z$ & 0&0&0&$\g$ & & 0&0&0&$\g$ & \texttt{refl}
\end{tabular}\hfill\vrule\hspace{6pt}%
\begin{minipage}[c]{0.32\linewidth}\raggedright\footnotesize
one clause per row,\\ for each $P\in\{\id,X,Y,Z\}$:\\
\texttt{ZZ\(\simeq\)CX-IZ-CX (}$P$\texttt{ \(\dblcolon\) \_ \(\dblcolon\) [])}\\
\texttt{\ \ = refl}
\end{minipage}\\[4pt]
\raggedright\small
\textbf{(3) Evaluate}: equal tables give equal Hamiltonians for every valuation $\nu$ and every $k$:\\[1pt]
\texttt{\footnotesize sound : p \(\simeq\) q \(\to\) \(\forall\) k \(\nu\) Q \(\to\) \(\llbracket\) p \(\rrbracket\)\^{} k \(\nu\) Q \(\equiv\) \(\llbracket\) q \(\rrbracket\)\^{} k \(\nu\) Q}\tabularnewline
\bottomrule
\end{tabular}
\caption{Agda proof of $ZZ(\gamma)\simeq CX\,(\id\otimes Z(\gamma))\,CX^\dagger$: expand, compare per Pauli string, evaluate.}
\Description{Three steps. First, both programs are expanded into Pauli strings and both equal gamma times Z tensor Z, the second via conjugation by CX mapping IZ to ZZ. Second, the two tables of Pauli coefficients side by side with an equals sign, one row per first Pauli, zero everywhere except gamma on ZZ, each row closed by refl, with the clause pattern for each first Pauli. Third, the soundness lemma giving equality under every valuation and level.}
\label{fig:agda}
\end{figure}

We mechanized this argument for a fragment of \VH in Agda \cite{agda} \codeagda{}, where \texttt{ZZ} and \texttt{CX-IZ-CX} are the two programs.

$\llbracket p\rrbracket$ is the Hamiltonian of $p$ as a list of Pauli strings, weighted by affine forms in which $\gamma$ is a symbol for $\nu(\gamma)/2^k$. A string \texttt{Q} lists one Pauli per qubit, so \texttt{Z\,\(\dblcolon\)\,\_\,\(\dblcolon\)\,[]} is $Z$ on the first qubit and any Pauli on the second. \texttt{coef\,\(\llbracket\)p\(\rrbracket\)\,Q} adds the weights on \texttt{Q}, and \texttt{\(\llbracket\)p\(\rrbracket\)\^{}\,k\,\(\nu\)\,Q} evaluates that sum at valuation $\nu$; the result corresponds to the coefficient of \texttt{Q} in $\llbracket p\rrbracket_H^\nu$.

In the proof (Fig.~\ref{fig:agda}), $\llbracket\cdot\rrbracket$ expands both programs, rewriting strings by conjugation tables for $H$, $S$, $CX$, $SWAP$, and appending lists for $\parallel$; \texttt{refl} checks all the coefficients, in four clauses, one per first Pauli, succeeding when Agda reduces both sides to identical forms; and \texttt{sound} extends this to every $\nu$ and $k$.

The development (App.~\ref{app:agda}) covers qubit-sized direct sums, Clifford transformations, conjugation, composition, tensor products and $\parallel$, but not compilation. The Pauli basis determines a Hamiltonian uniquely. Equality of all Pauli coefficients therefore establishes Hamiltonian equivalence.

\begin{table}[H]
\caption{The Pauli-sum semantics $\llbracket\cdot\rrbracket$ of the Agda fragment, the rules that \texttt{refl} evaluates.}
\label{tab:rules}
\centering\small
\begin{tabular}{@{}l>{\raggedright\arraybackslash}p{0.8\linewidth}@{}}
\toprule
\multicolumn{2}{@{}l}{\textbf{Construct}}\\\midrule
fixed clock & $\llbracket\pm\delta_k\rrbracket=\pm2^{-k}$ \quad (on the empty string)\\
open clock & $\llbracket\pm\delta_k^\alpha\rrbracket=\pm\alpha$, \quad where $\alpha$ stands for $\nu(\alpha)/2^k$\\
direct sum & $\llbracket d_0\oplus d_1\rrbracket=\id\otimes\tfrac12(\llbracket d_0\rrbracket+\llbracket d_1\rrbracket)+Z\otimes\tfrac12(\llbracket d_0\rrbracket-\llbracket d_1\rrbracket)$\\
conjugation & $\llbracket uau^\dagger\rrbracket=\sum_P c_P\,\sigma_u(P)\,u(P)$, \quad where $uPu^\dagger=\sigma_u(P)\,u(P)$, $\sigma_u(P)=\pm1$\\
parallel & $\llbracket p\parallel q\rrbracket=\llbracket p\rrbracket+\llbracket q\rrbracket$ \quad (lists appended)\\
\midrule
\multicolumn{2}{@{}l}{\textbf{Conjugation tables} $P\mapsto\sigma_u(P)\,u(P)$}\\\midrule
$H$ & $\id\mapsto\id$,\ \ $X\mapsto Z$,\ \ $Y\mapsto -Y$,\ \ $Z\mapsto X$\\
$S$ & $\id\mapsto\id$,\ \ $X\mapsto Y$,\ \ $Y\mapsto -X$,\ \ $Z\mapsto Z$\\
$CX$ (control first) & $\id\id\mapsto\id\id$,\ \ $\id X\mapsto\id X$,\ \ $\id Y\mapsto ZY$,\ \ $\id Z\mapsto ZZ$,\ \ $X\id\mapsto XX$,\ \ $XX\mapsto X\id$,\ \ $XY\mapsto YZ$,\ \ $XZ\mapsto -YY$,\ \ $Y\id\mapsto YX$,\ \ $YX\mapsto Y\id$,\ \ $YY\mapsto -XZ$,\ \ $YZ\mapsto XY$,\ \ $Z\id\mapsto Z\id$,\ \ $ZX\mapsto ZX$,\ \ $ZY\mapsto\id Y$,\ \ $ZZ\mapsto\id Z$\\
$SWAP$ & $PQ\mapsto QP$\\
$u\cdot v$,\ $u\otimes v$ & apply $v$ then $u$; apply $u$ and $v$ to the two parts of the string; signs multiply\\
\bottomrule
\end{tabular}
\end{table}

Table~\ref{tab:rules} lists the rules defining the Agda interpretation. The direct-sum rule follows from the identity $\mathrm{diag}(d_0,d_1)=\id\otimes\frac{d_0+d_1}2+Z\otimes\frac{d_0-d_1}2$. The conjugation tables were checked numerically against matrices; formally establishing agreement with \Ham's matrix semantics remains future work.

\FloatBarrier
\section{Running Programs with Open Clocks}\label{sec:running}
Fix a starting state $|\psi_0\rangle$, a target Hamiltonian $H_{\mathrm{tar}}$, and an ansatz of \VH blocks. Each valuation gives a state $|\psi(\nu)\rangle$ and an energy, and VQE seeks a valuation of lowest energy over an admissible set $\mathcal A$:
\[
E(\nu)=\langle\psi(\nu)|H_{\mathrm{tar}}|\psi(\nu)\rangle,\qquad \nu^\star\in\operatorname*{arg\,min}_{\nu\in\mathcal A}E(\nu).
\]
Here $\mathcal A$ is a grid of values of the clocks $\delta_r^\alpha$ at dyadic precision $r$. For QAOA it is the half range $\nu(\alpha)\in\{0,\dots,2^{r-1}-1\}$, which is one period of each block: at clock value $1/2$ both blocks return to the identity, up to a global phase. For the noncommuting block below it is the full range $\nu(\alpha)\in\{0,\dots,2^r-1\}$. The program fixes the motions; the optimizer chooses where to stop them.

Once its clocks are set, a \VH program runs: our Haskell prototype instantiates each valuation as a \Ham process, compiles it to OpenQASM at Trotter level $n$, and simulates the circuit's statevector \cite{vhamiltoncode} (App.~\ref{app:repro}), giving the compiled energy $E_n(\nu)$, without sampling noise. For blocks whose terms commute, such as QAOA's, compilation is exact and $E_n=E$.

\begin{figure}[t]
\centering
\input{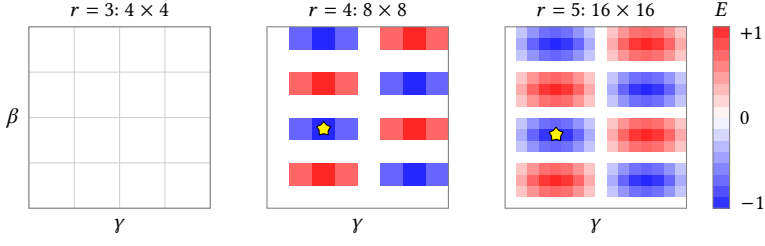}
\caption{QAOA energy $E=\sin(4\pi\gamma)\sin(8\pi\beta)$ on the one-period grid, $r=3,4,5$; optimum starred.}
\Description{Three heatmaps of the QAOA energy at r=3, 4 and 5: the first is uniformly zero; the others show alternating positive and negative regions, with the minimum energy minus one starred at clock values one eighth and three sixteenths.}
\label{fig:qaoa}
\end{figure}

\textbf{QAOA.} For MaxCut on one edge, $H_{\mathrm{tar}}=Z\otimes Z$. Its ground states $|01\rangle,|10\rangle$ cut the edge, with energy $-1$. Writing $\gamma$ and $\beta$ for the clock values $\nu(\gamma)/2^r$ and $\nu(\beta)/2^r$, the ansatz of Fig.~\ref{fig:code}a has energy \codecite{qaoa_one_edge}{qaoa}
\[
E=\sin(4\pi\gamma)\,\sin(8\pi\beta).
\] Both blocks have period $1/2$, so the one-period grid covers the whole landscape (Fig.~\ref{fig:qaoa}). At $r=3$ all 16 energies are exactly $0$: every grid point hits a zero of the $\beta$-dependence, so the coarse grid sees a flat landscape.

At $r=4$, exhaustive search over 64 valuations finds the clock values $(1/8,3/16)$, with $E=-1$ exactly, preparing a Bell state. At $\gamma=1/8$ the cost block is exactly $ZZ_3$ of Sec.~\ref{sec:hamilton} (Fig.~\ref{fig:fixed}c), and $\beta=3/16$ is the mixing time that turns its phases into the cut: the optimizer has rediscovered a hand-written program, and since both are \Ham programs, the identity is an equation between programs.

At $r=5$ the same point returns as $\nu^\star=(4,6)$, since $q/2^r=2q/2^{r+1}$, after 256 energy evaluations rather than 64: past the needed precision, extra bits only cost. Precision is thus a resource the programmer chooses; a coarse grid can be blind to the optimum, and a fine one pays for resolution it does not need.

\textbf{A noncommuting block.} Unlike the commuting terms within each QAOA block, $Z(\alpha)\parallel X(\beta)$ of Fig.~\ref{fig:code}b is not compiled exactly by our compiler, which Trotterizes it; once the clocks are set, the motion could be synthesized exactly as a single rotation, but that is not the strategy the compiler implements.

Running $Z$ and $X$ at once turns the Bloch sphere about one axis between them, tilted by the clocks, so a state starting at the north pole $|0\rangle$ moves on a circle around that axis. Take $H_{\mathrm{tar}}=X+Y+Z$, whose energy at a point $(x,y,z)$ of the sphere is $x+y+z$. Its ground state is the point opposite $(1,1,1)$, with energy $-\sqrt3\approx-1.732$.

With both clocks forward the axis lies between $+X$ and $+Z$, and no circle through the north pole around it reaches that point: the program's best is $-\sqrt2\approx-1.414$, on the $r=4$ grid and over all forward times \codecite{trotter_vqe}{trotter}.

As in Fig.~\ref{fig:fixed}d, level $n$ alternates $2^n$ turns about each axis, each $1/2^n$ of the full turn. At level $1$, with $A$ and $B$ the full turns, the circuit runs
\[
\sqrt{A}\,\sqrt{B}\,\sqrt{A}\,\sqrt{B}.
\] If the turns commuted, they would recombine into the program's single turn; here they do not, so the zigzag can leave the circle. The optimizer, which only sees circuit energies, then minimizes the zigzag's energy: on the $r=4$ grid it reports $-1.714$ at $\nu=(2,11)$, below anything the program reaches, while the program's own energy there is $-1.016$ (Table~\ref{tab:trotter}, which gives, for each level $n$, the compiled optimum over the 256 valuations, the program's own energy there, and the mean compilation error).

The compiled circuits form a different, richer family, and the search exploits its error. Coarse compilation is thus not merely imprecise: it changes the landscape being searched, which approaches the program's own only as $n$ grows, and a compiled optimum at a coarse level need not be an optimum of the program.

Because the program keeps its exact meaning, the discrepancy can be checked: at levels $2$, $3$ and $4$ the same valuation gives $-1.357$, $-1.148$ and $-1.072$, closing in on $-1.016$, and from level $2$ on the search returns the program's optimum. To make the low-energy zigzag the intended computation instead, one writes its four half-turns as sequential ansatz blocks rather than as the compilation of one simultaneous process.

\begin{table}[t]
\caption{Exhaustive search of $Z(\alpha)\parallel X(\beta)$ on the $r=4$ grid, by Trotter level $n$.}
\label{tab:trotter}
\centering\small
\begin{tabular}{@{}cccc@{}}
\toprule
$n$ & $\min_\nu E_n(\nu)$ & $E(\nu_n^\star)$ & $\operatorname{mean}_\nu|E_n-E|$\\
\midrule
0 & $-1.414$ & $-1.414$ & 1.04\\
1 & $\mathbf{-1.714}$ & $\mathbf{-1.016}$ & 0.64\\
2 & $-1.414$ & $-1.414$ & 0.53\\
3 & $-1.414$ & $-1.414$ & 0.24\\
4 & $-1.414$ & $-1.414$ & 0.10\\
\bottomrule
\end{tabular}
\end{table}

These examples are small enough to analyze by hand; the next section turns to a chemistry VQE \codecite{hamilton_vqe}{vqe}.

\section{Chemistry with Open Clocks}\label{sec:chem}
The examples so far are small enough to read off by hand. We now write a real variational eigensolver in \VH: a two-qubit molecular Hamiltonian as the target, and a three-clock ansatz that reaches its ground state. The point is not the chemistry but what the language makes visible: the target is a \Ham program at a chosen precision, the ansatz is a \VH program whose prepared states have a closed form, and two of its blocks can be fused by an equation between programs.

Proofs of the results in this section and the next are in App.~\ref{app:proofs}.

\subsection{The Target as a \Ham Program}\label{sec:target}
The target is the upper $4\times4$ block of the DMANH$^+$ Hamiltonian from the \Ham chemistry benchmark \cite{richerme2023}, which models a shared proton in a hydrogen bond. We symmetrize the reported block, removing a tiny asymmetry in its decimal data, and expand it in two-qubit Pauli strings (Table~\ref{tab:pauli}). Its ground energy is $E_0(H_{\mathrm{phys}})=-12.162810\,$mHa.

To put the coefficients on the scale of the clocks, we multiply by $s=0.019212480427$, which makes the largest non-identity coefficient $1/4$, and write
\[
\bar H=sH_{\mathrm{phys}}=\sum_\ell\bar c_\ell P_\ell .
\] At \emph{Hamiltonian precision} $k$ each coefficient is rounded to the nearest multiple of $2^{-k}$:
\[
c^{(k)}_\ell=2^{-k}\operatorname{round}\big(2^k\bar c_\ell\big),\qquad H_k=\sum_\ell c^{(k)}_\ell P_\ell .
\]
Each $c^{(k)}_\ell$ is a dyadic number, a finite sum of clocks $\pm\delta_j$ with $j\le k$ by its binary expansion, and each $P_\ell$ is a diagonal Pauli string conjugated by Cliffords, so $H_k$ is an ordinary \Ham program with fixed clocks. Energies are reported in mHa, dividing by $s$. Two precisions now play different roles: $k$ fixes the clocks of the target, and $r$ the grid of the ansatz's open clocks.

\begin{table}[t]
\caption{Pauli expansion of the symmetrized DMANH$^+$ upper block, in mHa.}
\label{tab:pauli}
\centering\small
\begin{tabular}{@{}lr@{\qquad}lr@{\qquad}lr@{\qquad}lr@{\qquad}lr@{}}
\toprule
$\id\id$ & $9.418625$ & $Z\id$ & $13.012375$ & $\id Z$ & $9.557625$ & $\id X$ & $-7.222925$ & $ZZ$ & $5.691375$\\
$XX$ & $-3.877200$ & $YY$ & $-3.874800$ & $X\id$ & $0.779450$ & $ZX$ & $-0.313925$ & $XZ$ & $-0.061250$\\
\bottomrule
\end{tabular}
\end{table}

Two preprocessing steps deserve care, because dynamics and energy treat them differently. The identity coefficient is invisible to evolution but not to energy:

\begin{proposition}[The identity term]\label{prop:identity}
For Hermitian $H$, $c\in\mathbb R$ and every $t$, $e^{-it(H+c\id)}=e^{-itc}e^{-itH}$, while $\langle\psi|H+c\id|\psi\rangle=\langle\psi|H|\psi\rangle+c$ for every normalized $|\psi\rangle$.
\end{proposition}

\noindent So the compiler may drop $\id\id$ as a global phase, but the energy must keep it. Our prototype keeps both forms: the target with its identity term, for energies, and without it, for circuits. At $k=14$ the scaled identity coefficient $0.18095515$ rounds to $2965/2^{14}=0.18096924$. Scaling, in turn, changes units but not answers:

\begin{proposition}[Positive affine invariance]\label{prop:affine}
Let $\widetilde H=aH+b\id$ with $a>0$. For every valuation, $E_{\widetilde H}(\nu)=aE_H(\nu)+b$; hence $E_{\widetilde H}$ and $E_H$ have the same minimizing valuations, and $E_0(\widetilde H)=aE_0(H)+b$.
\end{proposition}

\noindent Rounding is different: rounding the non-identity coefficients changes the Hamiltonian in a direction, not just by a shift, and can move the optimum, which is why $k$ appears in the error calculus of Sec.~\ref{sec:errors}. Rounding the identity coefficient only shifts every energy by the same amount.

\subsection{The Ansatz as a \VH Program}\label{sec:ansatz}
The ansatz of Fig.~\ref{fig:stairs}a uses three blocks of Sec.~\ref{sec:examples}, each with its own clock at precision $r$:
\[
YI(\alpha_1),\qquad CY(\alpha_2),\qquad \overline{C}Y(\alpha_3),
\]
where $YI(\alpha_1)$ is the $SH\otimes\id$ conjugate of $\delta_r^{\alpha_1}\oplus\delta_r^{\alpha_1}\oplus(-\delta_r^{\alpha_1})\oplus(-\delta_r^{\alpha_1})$, $CY(\alpha_2)$ is the controlled clock of Sec.~\ref{sec:examples}, and $\overline{C}Y(\alpha_3)$ is the same clock behind the opposite control, the $\id\otimes SH$ conjugate of $0\oplus0\oplus\delta_r^{\alpha_3}\oplus(-\delta_r^{\alpha_3})$. Their Hamiltonians are
\[
\frac{\nu(\alpha_1)}{2^r}\,Y\otimes\id,\qquad
\frac{\nu(\alpha_2)}{2^r}\,|0\rangle\langle0|\otimes Y,\qquad
\frac{\nu(\alpha_3)}{2^r}\,|1\rangle\langle1|\otimes Y.
\] The first block splits the amplitude between the two branches of the first qubit, and the other two rotate the second qubit within each branch. Each block is a \Ham generator with its time left open, an instance of a general fact:

\begin{lemma}[Opening a program]\label{lem:open}
For a \Ham process $p$, a variable $\alpha$ and $r\ge0$, let $p^{\alpha}_{r}$ replace every clock $\pm\delta_k$ of $p$ by $\pm\delta_{k+r}^\alpha$. Then $\llbracket p^{\alpha}_{r}\rrbracket_H^\nu=\frac{\nu(\alpha)}{2^r}\llbracket p\rrbracket_H$ for every valuation $\nu$.
\end{lemma}

\noindent So running a \Ham program for an open time is itself a \VH program, which makes precise the remark of Sec.~\ref{sec:hamilton} that running $H$ for time $t$ equals running $tH$ for one unit. Here the three blocks open the \Ham generators $Y\otimes\id$, $|0\rangle\langle0|\otimes Y$ and $|1\rangle\langle1|\otimes Y$, whose clocks are $\pm\delta_0$ and $0$.

Write $\phi_j=2\pi\nu(\alpha_j)/2^r$ for the angle through which block $j$ turns. Running the blocks in order from $|00\rangle$ prepares
\[
|\psi(\nu)\rangle=\llbracket\overline{C}Y(\alpha_3)\rrbracket_U^\nu\,\llbracket CY(\alpha_2)\rrbracket_U^\nu\,\llbracket YI(\alpha_1)\rrbracket_U^\nu\,|00\rangle,
\]
and the syntax tells us what that state is:

\begin{lemma}[Closed-form ansatz state]\label{lem:closed}
In the basis $|00\rangle,|01\rangle,|10\rangle,|11\rangle$,
\[
|\psi(\nu)\rangle=\big(\cos\phi_1\cos\phi_2,\ \cos\phi_1\sin\phi_2,\ \sin\phi_1\cos\phi_3,\ \sin\phi_1\sin\phi_3\big)^{\mathsf T}.
\]
\end{lemma}

\noindent The amplitudes are real, and every real state is reached:

\begin{theorem}[Projective real-state expressibility]\label{thm:express}
Every normalized real two-qubit state equals, up to a global sign, the state of Lemma~\ref{lem:closed} for some angles $\phi_1,\phi_2,\phi_3\in[0,\pi)$.
\end{theorem}

\noindent The angles $[0,\pi)$ are the clock values $[0,1/2)$. The \emph{half-range grid} samples them, letting each $\nu(\alpha_j)$ range over $\{0,\dots,2^{r-1}-1\}$. For $YI(\alpha_1)$ this is one period, since its turn at clock value $1/2$ is $e^{-i\pi\,Y\otimes\id}=-\id$, a global phase. For the controlled clocks it is not, since
\[e^{-i\pi\,|0\rangle\langle0|\otimes Y}=-Z\otimes\id\]
is not a global phase; their period is $1$.

The half range nevertheless suffices for the family of states. By Theorem~\ref{thm:express} it reaches every real state up to sign, through symmetries that change several clocks at once (Lemma~\ref{lem:closedgrid} below), not because each clock repeats after $1/2$.

Every $H_k$ is real and symmetric, since its Pauli strings, including $YY$, are real matrices, so it has a real ground state. The continuous family therefore contains a ground state of every $H_k$, and any remaining error comes from the grid, not from the shape of the program.

This ansatz also compiles exactly. Each block is a single action, so there is nothing to interleave, and its unitary is computed exactly at every Trotter level.

Our prototype writes the controlled blocks instead in Pauli form, with each $\pm$ coefficient at precision $r+1$: $CY(\alpha_2)$ as $IY(\alpha_2)\parallel ZY(\alpha_2)$ and $\overline CY(\alpha_3)$ as $IY(\alpha_3)\parallel -ZY(\alpha_3)$. These are Hamiltonian-equivalent to the actions by the expansion rule of Table~\ref{tab:rules}:
\[
|0\rangle\langle 0|\otimes Y=\tfrac12(\id\otimes Y+Z\otimes Y),\qquad |1\rangle\langle 1|\otimes Y=\tfrac12(\id\otimes Y-Z\otimes Y).
\] Since
\(
(\id\otimes Y)(Z\otimes Y)=Z\otimes\id=(Z\otimes Y)(\id\otimes Y),
\)
the two terms commute, so each level-$n$ product $(e^{-iA/2^n}e^{-iB/2^n})^{2^n}$ equals $e^{-i(A+B)}$, and this form too is exact at every level. The Trotter level $n$ therefore plays no part in this ansatz, as the compiler confirms (App.~\ref{app:compiler}), which isolates the two precisions $k$ and $r$; Sec.~\ref{sec:scaling} turns $n$ back on.

\subsection{Fusing the Controlled Blocks}\label{sec:fusion}
Two of the blocks commute, and \VH can say so. $CY(\alpha_2)$ acts only on the $|0\rangle$ branch of the first qubit and $\overline{C}Y(\alpha_3)$ only on the $|1\rangle$ branch, so the product of their Hamiltonians is $0$ in either order. They can be merged into one action with two clocks, which turns both branches at once:
\[
\mathit{Mux}(\alpha_2,\alpha_3):=(\id\otimes SH)\big(\delta_r^{\alpha_2}\oplus(-\delta_r^{\alpha_2})\oplus\delta_r^{\alpha_3}\oplus(-\delta_r^{\alpha_3})\big)(\id\otimes SH)^\dagger .
\]
That $\mathit{Mux}(\alpha_2,\alpha_3)\simeq CY(\alpha_2)\parallel\overline CY(\alpha_3)$ is a Hamiltonian equivalence of the kind proved in Sec.~\ref{sec:equivalence}, and we have checked it in Agda the same way \codeagda: here conjugation moves the second Pauli, so the proof splits on both Paulis, with one \texttt{refl} clause for each of the 16 strings.

Agda also computes the fused block's Pauli form, $\frac12(\alpha_2+\alpha_3)$ on $\id Y$ and $\frac12(\alpha_2-\alpha_3)$ on $ZY$, with each $\alpha_j$ a symbol for $\nu(\alpha_j)/2^r$. The equivalence relates a \emph{parallel} composition, however, while the ansatz runs the blocks in \emph{sequence}. Commutation connects the two:

\begin{proposition}[Sequence and parallel agree on commuting blocks]\label{prop:seqpar}
If $\llbracket p\rrbracket_H^\nu$ and $\llbracket q\rrbracket_H^\nu$ commute, then $\llbracket q\rrbracket_U^\nu\,\llbracket p\rrbracket_U^\nu=\llbracket p\parallel q\rrbracket_U^\nu$.
\end{proposition}

\begin{corollary}[Block fusion]\label{cor:fusion}
For every valuation $\nu$,
\[
\llbracket\overline CY(\alpha_3)\rrbracket_U^\nu\,\llbracket CY(\alpha_2)\rrbracket_U^\nu=\llbracket\mathit{Mux}(\alpha_2,\alpha_3)\rrbracket_U^\nu ,
\]
so the ansatz $[\,YI(\alpha_1),\ \mathit{Mux}(\alpha_2,\alpha_3)\,]$ prepares the same state as the three-block ansatz for every valuation.
\end{corollary}

\noindent \Ham has no commutator construct, so commutation is not itself an equation the calculus states; it is a property of the denoted Hamiltonians.

In the Pauli basis it is again a finite check. Two Pauli strings either commute or anticommute, so the commutator of two Pauli sums is a Pauli sum whose coefficients are products of clock values, and here every pair of strings, $\id Y$ and $ZY$, commutes.

The rewrite pays off in the circuit. In Pauli form the two controlled blocks are four rotations, two of them on two qubits ($ZY$). Merging terms on the same string is itself a Hamiltonian equivalence, so the fused block is two rotations,
\[
e^{-i\theta_+\,\id\otimes Y}\quad\text{and}\quad e^{-i\theta_-\,Z\otimes Y},\qquad \theta_\pm=\frac{\pi\,(\nu(\alpha_2)\pm\nu(\alpha_3))}{2^r}:
\]
one two-qubit rotation instead of two. The optimizer is unaffected: it still sets $\alpha_2$ and $\alpha_3$ on the same grid and sees the same energies, because both forms prepare the same state.

Only $YI(\alpha_1)$ must keep its place. It does not commute with the others; the $ZY$ term gives
\[
[Y\otimes\id,\,|0\rangle\langle0|\otimes Y]=\tfrac12[Y\otimes\id,\,Z\otimes Y]=iX\otimes Y.
\] So $YI(\alpha_1)$ stays first, before the fused pair.

\subsection{Two Precisions, One Error Budget}\label{sec:errors}
Two finite choices now separate a computed energy from the physical one: the precision $k$ of the target's clocks and the precision $r$ of the ansatz's clocks. Because the language keeps both exact, the error splits exactly along them. Let $E^\star_{k,r}=\min_\nu E_{H_k}(\nu)$ be the best energy on the half-range grid $\nu(\alpha_j)\in\{0,\dots,2^{r-1}-1\}$, and define, in scaled units,
\[
\delta_H(k)=E_0(H_k)-E_0(\bar H),\qquad
\delta_r(k,r)=E^\star_{k,r}-E_0(H_k),\qquad
\delta_{\mathrm{tot}}(k,r)=E^\star_{k,r}-E_0(\bar H).
\]
$\delta_H$ is what rounding the target costs, and $\delta_r$ what the grid costs on the rounded target.

\begin{proposition}[Exact error decomposition]\label{prop:decomp}
$\delta_{\mathrm{tot}}(k,r)=\delta_H(k)+\delta_r(k,r)$, with $\delta_r(k,r)\ge0$; hence $|\delta_{\mathrm{tot}}(k,r)|\le|\delta_H(k)|+\delta_r(k,r)$.
\end{proposition}

\noindent The two parts can have opposite signs: rounding may push the ground energy down while the grid always pushes the computed energy up. A small total error can therefore come from cancellation, which is real but cannot be planned for.

We call $|\delta_H(k)|+\delta_r(k,r)$ the \emph{noncancelling budget} and choose precisions against it. Each part has an a priori bound that needs no search. For the target:

\begin{theorem}[Rounding the target]\label{thm:hround}
If $\bar H$ has $M$ Pauli coefficients, each rounded to the nearest multiple of $2^{-k}$, including those that round to zero, then
\(
|\delta_H(k)|\le\|H_k-\bar H\|\le M\,2^{-(k+1)}.
\)
\end{theorem}

\noindent Here $M=10$, counting the identity, and dividing by $s$ converts to mHa. Each extra bit of $k$ halves the bound, and computing it requires no diagonalization.

For the clocks we view the energy as a function of continuous clock values $t\in\mathbb R^3$,
\[
E_k(t)=\langle\psi(t)|H_k|\psi(t)\rangle,\qquad \phi_j=2\pi t_j\ \text{in Lemma~\ref{lem:closed}};
\] the grid samples $t_j=\nu(\alpha_j)/2^r$ in $[0,1/2)$. Rounding a continuous optimum to the grid can land on the far edge $t_j=1/2$, which the half-range grid omits. The program's symmetries take care of this:

\begin{lemma}[The closed grid]\label{lem:closedgrid}
Every point of $\big(2^{-r}\mathbb Z\cap[0,\tfrac12]\big)^3$ has the same energy, for every $H_k$, as a point of the half-range grid.
\end{lemma}

\begin{theorem}[Rounding the clocks]\label{thm:rround}
Let $G_1=Y\otimes\id$, $G_2=|0\rangle\langle0|\otimes Y$ and $G_3=|1\rangle\langle1|\otimes Y$ be the generators of the $m=3$ blocks, so that $\|G_j\|=1$, and let $\|H_k\|$ be the operator norm. Then
\[
\delta_r(k,r)\ \le\ 2\pi m\,\|H_k\|\,2^{-r}
\qquad\text{and}\qquad
\delta_r(k,r)\ \le\ 2\pi^2m^2\,\|H_k\|\,4^{-r}.
\]
\end{theorem}

\noindent The first bound holds wherever the energy is rounded; the second uses that the ground state is reached (Theorem~\ref{thm:express}), so the continuous optimum is a global minimum of $E_k$ on all of $\mathbb R^3$, where the gradient vanishes and only second-order terms remain. Neither needs $H_k$ diagonalized, since $\|H_k\|\le\sum_\ell|c^{(k)}_\ell|$. Each extra bit of $r$ thus divides the guaranteed gap by four. Finer grids also never hurt:

\begin{theorem}[Nested grids]\label{thm:nested}
For each $k$, $\delta_r(k,r+1)\le\delta_r(k,r)$, and $\delta_r(k,r)\to0$ as $r\to\infty$.
\end{theorem}

\noindent Monotonicity is the nesting $q/2^r=2q/2^{r+1}$ of Sec.~\ref{sec:valuations}: the coarse grid is part of the fine one. The limit follows from Theorem~\ref{thm:rround}.

\begin{figure}[t]
\centering
\begin{tikzpicture}
\begin{groupplot}[group style={group size=2 by 1, horizontal sep=1.3cm},
  width=0.38\linewidth, height=5cm, ymode=log, grid=major, grid style={gray!20},
  tick label style={font=\scriptsize}, label style={font=\footnotesize},
  legend style={font=\scriptsize, draw=none, fill=white, fill opacity=0.8, text opacity=1}, legend cell align=left]
\nextgroupplot[xlabel={Hamiltonian precision $k$}, ylabel={mHa}, xtick={4,6,...,14},
  title={\footnotesize (a) rounding the target}, ymin=1e-6, ymax=1e2, legend pos=south west]
\addplot[thick, violet, mark=*, mark size=1.6pt] coordinates {(3,5.437696e+00) (4,8.543236e-01) (5,2.828513e-01) (6,1.954327e-01) (7,2.390208e-01) (8,7.054425e-02) (9,6.278475e-02) (10,2.045404e-02) (11,7.058311e-03) (12,6.009903e-03) (13,3.298337e-03) (14,4.401121e-04)};
\addlegendentry{$|\delta_H(k)|$}
\addplot[thick, dashed, black!60] coordinates {(3,3.253094e+01) (4,1.626547e+01) (5,8.132734e+00) (6,4.066367e+00) (7,2.033184e+00) (8,1.016592e+00) (9,5.082959e-01) (10,2.541479e-01) (11,1.270740e-01) (12,6.353699e-02) (13,3.176849e-02) (14,1.588425e-02)};
\addlegendentry{bound $M2^{-(k+1)}/s$}
\nextgroupplot[xlabel={clock precision $r$}, xtick={2,...,8},
  title={\footnotesize (b) the grid of clock values}, ymin=1e-4, ymax=1e3,
  legend style={at={(1.04,0.5)}, anchor=west}]
\addplot[thick, black!35, mark=square*, mark size=1.6pt] coordinates {(2,5.038782e+00) (3,1.785688e+00) (4,7.461830e-01) (5,3.540787e-02) (6,1.638786e-02) (7,1.321748e-02) (8,3.467721e-03)};
\addlegendentry{$\delta_r(6,r)$}
\addplot[thick, black!60, mark=triangle*, mark size=1.6pt] coordinates {(2,4.711315e+00) (3,1.661539e+00) (4,7.955770e-01) (5,2.575992e-02) (6,1.183128e-02) (7,9.583139e-03) (8,1.228216e-03)};
\addlegendentry{$\delta_r(10,r)$}
\addplot[thick, violet, mark=*, mark size=1.6pt] coordinates {(2,4.704008e+00) (3,1.660587e+00) (4,7.936937e-01) (5,2.583635e-02) (6,1.192176e-02) (7,9.404619e-03) (8,1.187166e-03)};
\addlegendentry{$\delta_r(14,r)$}
\addplot[thick, dotted, black!60] coordinates {(2,1.866821e+02) (3,9.334103e+01) (4,4.667051e+01) (5,2.333526e+01) (6,1.166763e+01) (7,5.833814e+00) (8,2.916907e+00)};
\addlegendentry{first-order bound}
\addplot[thick, dashed, black!60] coordinates {(2,4.398592e+02) (3,1.099648e+02) (4,2.749120e+01) (5,6.872801e+00) (6,1.718200e+00) (7,4.295500e-01) (8,1.073875e-01)};
\addlegendentry{second-order bound}
\end{groupplot}
\end{tikzpicture}
\caption{Measured errors and their a priori bounds (mHa, log scale): (a) target rounding; (b) grid gap.}
\Description{Two log-scale line plots. Left: the absolute ground-energy shift from rounding the target falls from about 5 mHa at k=3 to 0.0004 mHa at k=14, not monotonically, always below a dashed bound that halves with each bit. Right: the grid gap for k=6, 10 and 14 falls from about 5 mHa at r=2 to about 0.001 mHa at r=8, nearly the same for all three k, below a dotted first-order bound and a dashed second-order bound that falls by four per bit.}
\label{fig:errors}
\end{figure}
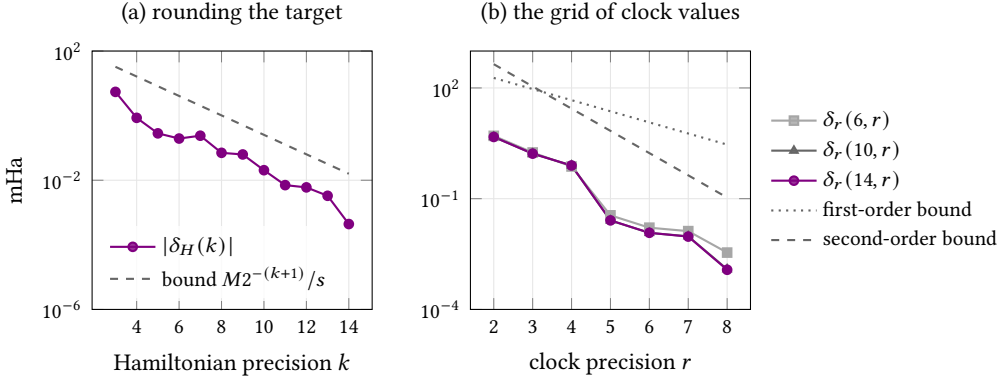

These bounds are guarantees, and on this benchmark they are loose: across $k\in\{6,10,14\}$ and $r=2,\dots,8$ the measured gap never exceeds $5\%$ of the second-order bound (Fig.~\ref{fig:errors}b, which draws the bounds of Theorem~\ref{thm:rround} at $k=14$; the curves for $k=10$ and $14$ nearly coincide).

For a small target there is a sharper estimate, built from the program's closed form. Diagonalize $H_k$, invert its real ground state through Lemma~\ref{lem:closed} as in the proof of Theorem~\ref{thm:express}, and round the angles to the grid once. The energy of that one valuation bounds the optimum:

\begin{proposition}[Constructive bound]\label{prop:construct}
For every valuation $\tilde\nu$ on the grid, $0\le\delta_r(k,r)\le E_{H_k}(\tilde\nu)-E_0(H_k)$.
\end{proposition}

\noindent Adding $|\delta_H(k)|$ gives a budget for each $(k,r)$ before any search. Table~\ref{tab:precision} picks, for each target accuracy, the smallest grid and then the smallest $k$ that meet it, scanning $k=3,\dots,16$ and $r=2,\dots,10$ \codecite{hamilton_vqe/05_pre_vqe_precision_selection}{precision}; its last column is the size of the half-range grid, $2^{3(r-1)}$. This procedure uses a ground-state eigenvector obtained by numerical diagonalization, so it is a check of the methodology on a target small enough to diagonalize, not a substitute for the eigensolver.

The 0.001\,mHa row predicts $r=10$ and the valuation $(223,238,155)$; the search of Sec.~\ref{sec:scaling} reaches that grid point after 53 energy queries rather than 134 million.

\begin{table}[t]
\caption{Precisions chosen before any search, from the constructive bound (mHa).}
\label{tab:precision}
\centering\small
\begin{tabular}{@{}rrrrrrr@{}}
\toprule
target & $k$ & $r$ & $|\delta_H(k)|$ & bound on $\delta_r$ & budget & grid\\
\midrule
$1$     & 5  & 4  & $0.282851$ & $0.575869$ & $0.858720$ & $512$\\
$0.1$   & 8  & 5  & $0.070544$ & $0.026276$ & $0.096820$ & $4{,}096$\\
$0.05$  & 10 & 5  & $0.020454$ & $0.025760$ & $0.046214$ & $4{,}096$\\
$0.01$  & 14 & 7  & $0.000440$ & $0.009405$ & $0.009845$ & $262{,}144$\\
$0.005$ & 13 & 8  & $0.003298$ & $0.001141$ & $0.004440$ & $2{,}097{,}152$\\
$0.001$ & 14 & 10 & $0.000440$ & $0.000119$ & $0.000559$ & $134{,}217{,}728$\\
\bottomrule
\end{tabular}
\end{table}

\textbf{Which precision matters.} The two axes behave differently (Fig.~\ref{fig:errors}). Rounding the target converges, but not monotonically: rounding can move the ground energy to either side of the physical value, so $|\delta_H(k)|$ rises from $k=6$ to $k=7$ even as its bound halves. The grid gap, by contrast, decreases at every step, as Theorem~\ref{thm:nested} requires, and is close for all three $k$. In the region studied $r$ dominates.

At $(k,r)=(10,3)$ exhaustive search returns $\nu=(2,0,1)$, one of four tied minimizers (at $\phi_1=\pi/2$ the $|0\rangle$ branch is empty, so $\alpha_2$ has no effect), with energy $-10.521725$\,mHa, against $E_0(H_{10})=-12.183264$\,mHa: the grid gap is $1.661539$\,mHa, while rounding the target costs only $\delta_H(10)=-0.020454$\,mHa \codecite{hamilton_vqe/06_k_r_error_separation}{separation}. Two more clock bits, at $(10,5)$, return $\nu=(7,7,5)$ and cut the gap to $0.025760$\,mHa.

There the signed total error is only $+0.005306$\,mHa, because the negative shift of the target partly cancels the positive gap, while the noncancelling budget is $0.046214$\,mHa.

Over the exhaustive sweep $k=3,\dots,12$, $r=2,\dots,6$, $(10,5)$ has the smallest total error and $(12,6)$ the smallest budget, $0.017954$\,mHa \codecite{hamilton_vqe/07_exhaustive_k_r_sweep}{sweep}. Precision is thus planned with the budget, and the cancellation is a bonus.

\section{Searching Many Clocks}\label{sec:scaling}
Two clocks can be searched exhaustively; $m$ clocks at precision $r$ cannot, since the grid has $2^{m(r-1)}$ points. We explore how the structure \VH makes visible can replace exhaustive search.

\begin{figure}[t]
\centering
\begin{tikzpicture}[x=1cm,y=1cm, font=\footnotesize]
  \begin{scope}[shift={(0.5,0)}]
    \draw[wire] (0,0) -- (4.0,0);
    \draw[wire] (0,-0.7) -- (4.0,-0.7);
    \node[left] at (0,0) {$|0\rangle$};
    \node[left] at (0,-0.7) {$|0\rangle$};
    \node[smallopen] at (0.6,0) {$\hole{\alpha_1}\,Y$};
    \draw[thick] (1.85,0) -- (1.85,-0.45);
    \filldraw[thick, fill=white] (1.85,0) circle (3pt);
    \node[smallopen] at (1.85,-0.7) {$\hole{\alpha_2}\,Y$};
    \draw[thick] (3.15,0) -- (3.15,-0.45);
    \fill (3.15,0) circle (3pt);
    \node[smallopen] at (3.15,-0.7) {$\hole{\alpha_3}\,Y$};
    \node[draw, thick, fill=white, minimum height=0.9cm, anchor=west] at (4.0,-0.35) {$\langle H_{\mathrm{mol}}\rangle$};
  \end{scope}
  \begin{scope}[shift={(6.95,0)}]
    \draw[wire] (0,0) -- (4.5,0);
    \draw[wire] (0,-0.7) -- (4.5,-0.7);
    \node[left] at (0,0) {$|0\rangle$};
    \node[left] at (0,-0.7) {$|0\rangle$};
    \draw[gray, dashed, rounded corners] (0.08,0.52) rectangle (2.62,-1.25);
    \node[font=\footnotesize, anchor=south east, inner sep=1pt] at (2.62,0.52) {$\times\,2^n$, clocks $/2^n$};
    \node[smallopen] at (0.68,0) {$\hole{\alpha_1}\,Y$};
    \node[draw, thick, dashed, fill=violet!8, minimum height=0.82cm, inner sep=2pt, font=\footnotesize] at (1.9,-0.35) {$\hole{\alpha_2}\,ZY$};
    \node[smallopen] at (3.45,-0.7) {$\hole{\alpha_3}\,Y$};
    \node[draw, thick, fill=white, minimum height=0.9cm, anchor=west] at (4.5,-0.35) {$\langle H_{\mathrm{mol}}\rangle$};
  \end{scope}
  \node[font=\footnotesize\bfseries] at (-0.35,0) {(a)};
  \node[font=\footnotesize\bfseries] at (6.15,0) {(b)};
  \node[font=\footnotesize\bfseries] at (-0.3,-1.55) {(c)};
  \foreach \i/\name in {0/{1 Every point},1/{2 Zoom in},2/{3 One knob},3/{4 Fit the wave},4/{5 Live start}} {
    \node[font=\footnotesize\bfseries] at (\i*2.75+1.1,-1.55) {\name};
    \draw[gray!50, rounded corners] (\i*2.75,-1.75) rectangle (\i*2.75+2.2,-3.3);
  }
  \begin{scope}[shift={(0.35,-3.15)}]
    \foreach \x in {0,...,6} \foreach \y in {0,...,5} \fill (\x*0.25,\y*0.24) circle (1.1pt);
  \end{scope}
  \begin{scope}[shift={(2.75+0.35,-3.15)}]
    \foreach \x in {0,1,2} \foreach \y in {0,1,2} \fill[gray] (\x*0.75,\y*0.6) circle (1.6pt);
    \draw[thick, violet] (0.75-0.3,0.6-0.3) rectangle (0.75+0.3,0.6+0.3);
    \foreach \x in {-1,0,1} \foreach \y in {-1,0,1} \fill[violet] (0.75+\x*0.18,0.6+\y*0.18) circle (1.1pt);
  \end{scope}
  \begin{scope}[shift={(2*2.75+1.1,-2.55)}]
    \fill (0,0) circle (2pt);
    \foreach \a in {0,180} \draw[flow, violet] (0,0) -- (\a:0.7);
    \foreach \a in {90,270} \draw[flow, violet] (0,0) -- (\a:0.55);
    \draw[gray, dashed] (-0.6,-0.5) -- (0.6,0.5);
  \end{scope}
  \begin{scope}[shift={(3*2.75+0.16,-2.52)}, x=0.30cm, y=0.4cm]
    \draw[thick, black!70] plot[domain=0:6.2832, samples=80, smooth]
      (\x, {0.15-0.35*cos(\x r)+0.25*sin(\x r)+0.3*cos(2*\x r)-0.2*sin(2*\x r)});
    \foreach \t in {0,1.2566,2.5133,3.7699,5.0265}
      \fill[violet] (\t, {0.15-0.35*cos(\t r)+0.25*sin(\t r)+0.3*cos(2*\t r)-0.2*sin(2*\t r)}) circle (2pt);
    \node[draw, circle, inner sep=1.4pt, fill=white] at (4.71,-0.40) {};
  \end{scope}
  \begin{scope}[shift={(4*2.75+0.3,-3.08)}]
    \fill[teal!60] (0,0) rectangle (0.55,0.75);
    \fill[violet!60] (0.75,0) rectangle (1.3,0.75);
    \draw[gray] (-0.1,0) -- (1.6,0);
    \node[font=\scriptsize] at (0.27,-0.18) {$|0\rangle$};
    \node[font=\scriptsize] at (1.02,-0.18) {$|1\rangle$};
    \node[font=\scriptsize] at (0.65,0.97) {$|{+}{+}\rangle$};
  \end{scope}
  \foreach \i/\lab in {0/{262{,}144},1/{315},2/{189},3/{\QAnalytic},4/{38}}
    \node[font=\footnotesize, anchor=north] at (\i*2.75+1.1,-3.35) {\lab\ queries};
  \begin{scope}[x=1cm, y=1cm, shift={(0,-4.05)}]
    \node[font=\footnotesize\bfseries] at (-0.3,0) {(d)};
    \foreach \x/\name in {0/{Schedules}, 4.45/{Error (\textemdash) and regret (-\,-)}, 8.9/{Cost to the optimum}} {
      \node[font=\footnotesize\bfseries] at (\x+2.0,0) {\name};
      \draw[gray!50, rounded corners] (\x,-0.2) rectangle (\x+4.0,-2.45);
    }
    \foreach \row/\y/\lab in {0/-0.8/{canon}, 1/-1.45/{sym}} {
      \node[font=\scriptsize, anchor=west] at (0.08,\y) {\lab};
      \foreach \rep in {0,1,2} {
        \foreach \k in {1,2} {
          \pgfmathtruncatemacro{\kk}{(\row==1 && \rep==1) ? 3-\k : \k}
          \node[draw=violet!70, fill=violet!8, inner sep=0pt, minimum size=0.21cm, font=\tiny]
            at (0.98+\rep*0.66+\k*0.25,\y) {\kk};
        }
      }
      \node[font=\scriptsize] at (3.18,\y) {$\cdots$};
      \node[draw=black!60, fill=white, inner sep=0pt, minimum size=0.21cm, font=\tiny] at (3.62,\y) {3};
    }
    \draw[decorate, decoration={brace, mirror, amplitude=3pt}] (1.1,-1.65) -- (3.35,-1.65);
    \node[font=\scriptsize] at (2.22,-2.15) {$2^n$ repetitions};
    \node[font=\scriptsize] at (3.62,-2.15) {once};
    \begin{scope}[shift={(5.05,-2.05)}]
      \draw[gray] (0,-0.25) -- (2.85,-0.25);
      \draw[gray] (0,-0.25) -- (0,-0.12);
      \draw[gray] (0,-0.04) -- (0,1.55);
      \draw[gray] (-0.06,-0.14) -- (0.06,-0.10);
      \draw[gray] (-0.06,-0.06) -- (0.06,-0.02);
      \foreach \n in {0,...,5} \node[font=\tiny, below] at (\n*0.55,-0.25) {\n};
      \node[font=\scriptsize] at (3.05,-0.43) {$n$};
      \foreach \v/\lab in {-0.25/{0}, 0.31/{.01}, 0.93/{1}, 1.55/{100}} {
        \draw[gray] (-0.05,\v) -- (0,\v);
        \node[font=\tiny, left] at (0,\v) {\lab};
      }
      \draw[thick, black!45] plot coordinates {(0,1.029) (0.55,0.913) (1.10,0.809) (1.65,0.710) (2.20,0.615) (2.75,0.520)};
      \draw[thick, violet] plot coordinates {(0,1.029) (0.55,0.977) (1.10,0.737) (1.65,0.538) (2.20,0.348) (2.75,0.161)};
      \draw[thick, dashed, black!45] plot coordinates {(0,1.401) (0.55,0.650) (1.10,0.388) (1.65,0.388) (2.20,-0.25) (2.75,-0.25)};
      \draw[thick, dashed, violet] plot coordinates {(0,1.401) (0.55,0.571) (1.10,-0.25) (1.65,-0.25) (2.20,-0.25) (2.75,-0.25)};
      \node[star, star points=5, fill=black!60, inner sep=1.7pt] at (2.20,-0.25) {};
      \node[star, star points=5, fill=violet, inner sep=1.7pt] at (1.10,-0.25) {};
      \node[font=\tiny, black!60, anchor=west] at (2.9,0.52) {canon};
      \node[font=\tiny, violet, anchor=west] at (2.9,0.16) {sym};
      \node[font=\tiny, anchor=west, inner sep=1pt] at (0.06,1.55) {mHa};
    \end{scope}
    \begin{scope}[shift={(9.35,-2.15)}]
      \draw[gray] (-0.1,0) -- (3.2,0);
      \foreach \i/\h/\lab/\name/\col in {0/1.30/{5{,}379}/{cold, $n{=}4$}/black!35, 1/0.413/{1{,}709}/{climb}/violet!45, 2/0.210/{867}/{climb, sym}/violet} {
        \fill[\col] (\i*1.05+0.15,0) rectangle (\i*1.05+0.85,\h);
        \node[font=\tiny, anchor=south] at (\i*1.05+0.5,\h) {\lab};
        \node[font=\tiny, anchor=north] at (\i*1.05+0.5,0) {\name};
      }
    \end{scope}
  \end{scope}
\end{tikzpicture}
\caption{Two three-clock ansatzes and their optimization costs: (a) commuting-block ansatz; (b) a noncommuting pair compiled by Trotterization; (c) five search strategies and their energy-query counts at $(k,r)=(14,7)$; (d) Trotter schedules, errors, and atom costs.}
\Description{(a) A two-qubit circuit with three open rotation clocks: a Y rotation and two controlled Y rotations. (b) The same three clocks with one Pauli generator each, the noncommuting pair repeated at Trotter level n and the free clock once. (c) Five pictograms of search strategies with their energy queries: 262,144; 315; 189; 128; 38. (d) Canonical and symmetric schedules, the error near the optimum and the regret against level n on a log scale, and atom applications to reach the optimum: 5,379; 1,709; 867.}
\label{fig:stairs}
\end{figure}
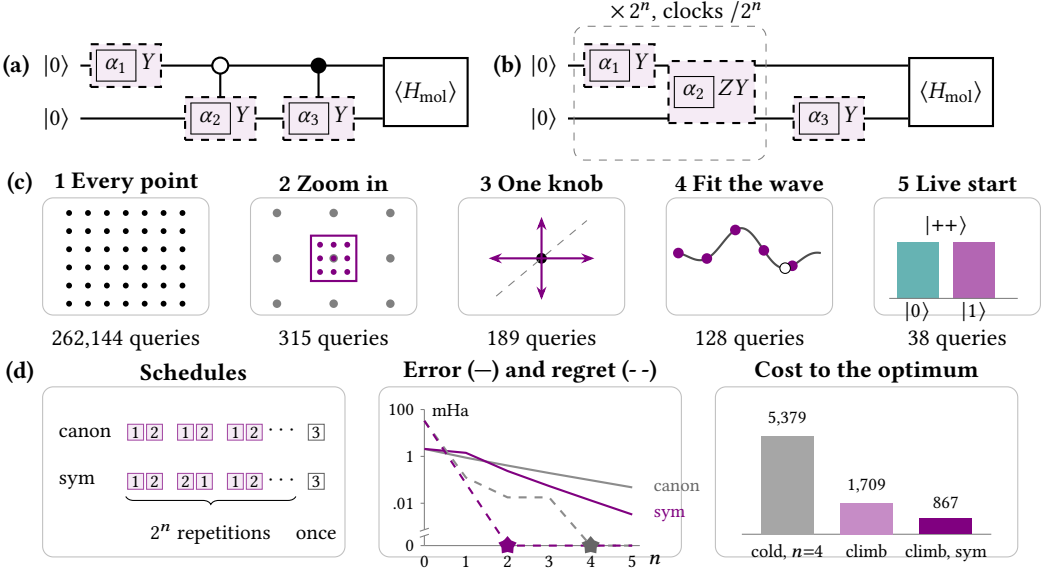

We use the target $H_{14}$ of Sec.~\ref{sec:target}, with Pauli coefficients rounded to $k=14$ dyadic bits, and two ans\"atze. The first (Fig.~\ref{fig:stairs}a) is the ansatz of Sec.~\ref{sec:ansatz}: one $Y$ rotation followed by two controlled rotations, one on each branch of the first qubit (open control: on $|0\rangle$; filled: on $|1\rangle$). At $r=7$ its grid has $262{,}144$ valuations.

Each step below removes one source of cost; Fig.~\ref{fig:stairs}c counts the energy queries our prototype made \codecite{hamilton_vqe/08_search_method_comparison}{search}, and steps 2--5 all return the exhaustive optimum $\nu=(28,30,19)$.

\subsection{Five Steps}\label{sec:steps}
\textbf{1 $\to$ 2, zoom instead of restarting.} Dyadic grids nest, $q/2^r=2q/2^{r+1}$: a time found at coarse precision is still on every finer grid. So search coarsely, keep a few promising valuations, and refine only around them, as one zooms into a map instead of reading it at full resolution everywhere. The QAOA run of Sec.~\ref{sec:running} shows the risk: its $r=3$ grid is flat (Fig.~\ref{fig:qaoa}).

\begin{proposition}[Beam refinement]\label{prop:beam}
Search the grid at precision $r_0$ exhaustively, keep the best $B$ valuations, and at each finer precision double them, $q\mapsto2q$, and evaluate all $3^m$ valuations within one grid step in every clock, keeping the best $B$. Reaching precision $r$ costs at most
\[
2^{m(r_0-1)}+B\,3^m\,(r-r_0)
\]
energy queries.
\end{proposition}

\noindent Our prototype uses $r_0=3$ and $B=3$; at $(k,r)=(14,7)$ the bound is $64+324=388$, and repeated valuations are cached, so the run makes $315$ queries. This count depends on how ties are broken; see the end of Sec.~\ref{sec:measured}.

\textbf{2 $\to$ 3, turn one knob at a time.} Comparing all neighbours of a point costs $3^m$ queries per round; moving one clock at a time costs $2m$. Like tuning one string of a guitar while holding the others, this is linear in $m$ per round, but it can stall when clocks interact strongly.

\begin{proposition}[Coordinate refinement]\label{prop:coord}
If at precision $\ell$ each of the $B$ kept valuations takes $L_\ell$ rounds of one-clock moves $\nu(\alpha_j)\mapsto\nu(\alpha_j)\pm1$, refining from $r_0$ to $r$ costs at most
\[
2^{m(r_0-1)}+2Bm\sum_{\ell=r_0+1}^{r}L_\ell
\]
energy queries.
\end{proposition}

\noindent The bound trades $3^m$ for $2m$ but multiplies by the number of rounds, and it carries no guarantee of finding the optimum: a lower valuation reachable only by moving two clocks together is invisible to one-clock moves.

\textbf{3 $\to$ 4, the syntax tells you the shape of the curve.} Hold all clocks but one fixed. If that clock drives a single block $e^{-2\pi i t G}$, the energy is a short Fourier series in $t$ whose frequencies are the eigenvalue gaps of $G$ \cite{wierichs2022}. In \VH those gaps are \emph{visible in the program}: conjugation preserves spectra, so the eigenvalues of $u\,d\,u^\dagger$ are the diagonal entries of $d$.

The $Y$ rotation has entries $\pm\delta_r^\alpha$ (one frequency, three unknowns); the controlled rotations have $\{0,\pm\delta_r^\alpha\}$ (two frequencies, five unknowns). A handful of queries therefore pins down the whole curve, whose minimum is found classically and snapped to the grid \cite{ostaszewski2021,nakanishi2020}, the way a few points determine a circle. The cost per clock no longer depends on the precision $r$; of the remaining \QAnalytic\ queries, half are spent on the coarse seed.

The following fact is known \cite{wierichs2022,nakanishi2020,ostaszewski2021}. We state it in \VH's terms and prove it in App.~\ref{app:proofs} for completeness. What is new here is where the spectrum comes from: it is read off the program's syntax, not supplied separately.

\begin{theorem}[The shape of one clock's curve \cite{wierichs2022}]\label{thm:spectral}
Fix every clock but $\alpha_j$, whose block runs $e^{-2\pi itG}$, and let $G=\sum_{\lambda}\lambda\Pi_\lambda$ be its spectral decomposition. With $\phi=2\pi t$, the energy is
\[
f(\phi)=\sum_{\lambda,\mu}e^{i(\lambda-\mu)\phi}\,\langle\eta|\Pi_\lambda A\Pi_\mu|\eta\rangle
=c+\sum_{\omega\in\Omega}\big(a_\omega\cos\omega\phi+b_\omega\sin\omega\phi\big),
\]
where $|\eta\rangle$ is the state before the block, $A$ is $H_k$ conjugated by the blocks after it, and $\Omega$ is the set of positive gaps $\lambda-\mu$ in the spectrum of $G$. So $f$ is determined by $2|\Omega|+1$ real numbers.
\end{theorem}

\noindent For the ansatz of Sec.~\ref{sec:ansatz}, $Y\otimes\id$ has spectrum $\{\pm1\}$, so $\Omega=\{2\}$, and the controlled generators have spectrum $\{-1,0,1\}$, so $\Omega=\{1,2\}$: three and five unknowns. Each curve is therefore fixed by energies at a few clock values, and these can be chosen on the grid once and for all.

\begin{corollary}[Constant queries per clock]\label{cor:samples}
For $r\ge4$, sampling $\phi\in\{0,\frac\pi4,\frac\pi2\}$ for $YI(\alpha_1)$, that is $\nu(\alpha_1)\in\{0,2^{r-3},2^{r-2}\}$, and $\phi\in\{0,\frac\pi4,\frac\pi2,\frac{3\pi}4,\frac{7\pi}8\}$ for each controlled clock determines its curve exactly. For $YI(\alpha_1)$,
\[
c=\frac{f(0)+f(\pi/2)}2,\qquad a_2=\frac{f(0)-f(\pi/2)}2,\qquad b_2=f(\tfrac\pi4)-c.
\] The number of samples is independent of $r$.
\end{corollary}

\noindent The fitted curve is then minimized classically, without further queries. A trigonometric polynomial with largest frequency $2$ has at most two local minima per period, and the best grid value of the clock lies within one step of one of them or of an end of the axis, so a fixed set of candidates suffices.

Two implementations use this differently. The NumPy reference minimizes the fitted curve exactly over the grid, up to floating point, by evaluating it at all $2^{r-1}$ values of the clock: classical work that grows with $r$, but no queries. The prototype locates the curve's local minima by a fixed scan with golden-section refinement and compares a fixed set of nearby grid values, so its classical work per fit does not grow with $r$. That numerical step is not proved to find every local minimum.

At the three settings both ran, $(k,r)=(10,5)$, $(14,7)$ and $(14,10)$, the two return the same valuations with the same query counts. Over the 66 fits of the compiler-backed runs, the fitted curve predicts the energy at its chosen point to within $3\times10^{-16}$ in scaled units, as expected of an exact model. Seeding as before gives:

\begin{proposition}[Fitted search from a coarse seed]\label{prop:analyticbeam}
Seeding with an exhaustive search at precision $r_0$, lifting the best $B$ valuations directly to precision $r$ by $q\mapsto2^{r-r_0}q$, and running $s$ sweeps of one-clock fits costs $O\big(2^{m(r_0-1)}+B\,m\,D\,s\big)$ energy queries, where $D$ bounds the number of frequencies per clock. No factor depends on $r$.
\end{proposition}

\textbf{4 $\to$ 5, start where every branch is alive.} Starting from $\nu=0$ leaves the first qubit in $|0\rangle$, so the clock behind the filled control acts on an empty branch and its curve is flat. Setting all three clock times to $1/8$ prepares $|{+}{+}\rangle$, in which both branches are populated (Fig.~\ref{fig:stairs}c), so no clock starts on an empty branch and coarse seeding can be dropped. What remains costs $O(m\cdot s\cdot D)$ queries for $s$ sweeps and at most $D$ frequencies per clock. Across precisions $r=4,\dots,10$ the grid grows from $512$ to $1.3\times10^8$ points while the queries grow from $18$ to $53$, ending $1.2\times10^{-4}$\,mHa above the dyadic target's ground energy \codecite{hamilton_vqe/12_analytic_direct_precision_scaling}{scaling}.

\begin{lemma}[A balanced start]\label{lem:seed}
For $r\ge3$, the valuation $\nu_0(\alpha_j)=2^{r-3}$ for every $j$ gives $\phi_j=\pi/4$ and prepares $|{+}{+}\rangle$, and among all states of the ansatz it maximizes the smallest of the four computational-basis probabilities, which is $1/4$.
\end{lemma}

\noindent The seed costs no queries to find, only $m$ assignments of the number $2^{r-3}$. Because it populates every branch, it rules out a flat curve caused by an empty control branch. A curve could still be flat for other reasons, for instance if a generator commuted with the observable it is measured against; on this benchmark, all three curves from the seed vary, by $0.29$ to $0.56$ in scaled units, for $k\in\{3,8,14\}$ and $r\in\{4,7,10\}$. Lemma~\ref{lem:seed} also does not say the seed is close to the optimum in energy. The search from it is counted exactly:

\begin{theorem}[Queries of the fitted search]\label{thm:direct}
Starting from $\nu_0$, a run of $s$ sweeps, each fitting every clock once by Corollary~\ref{cor:samples} and then querying the chosen valuation, makes at most
\[
1+s\sum_j\big(2|\Omega_j|+2\big)
\]
energy queries; for this ansatz, at most $1+16s$. In general this is $O(m\,D\,s)$, with no factor that depends on $r$.
\end{theorem}

\noindent The count is conditional on $s$, and $s$ is not bounded by a constant. What exact one-clock minimization does guarantee is that the search stops:

\begin{theorem}[Termination]\label{thm:terminate}
Suppose each step sets one clock to a value minimizing the energy along that clock on the grid, keeping the current value when it is among the minimizers, and the search stops after a sweep that changes no clock. Then the energy never increases, every change strictly lowers it, and the search stops after finitely many sweeps at a valuation $\nu^\ast$ that no single-clock change can improve.
\end{theorem}

\begin{corollary}[A conditional bound]\label{cor:conditional}
Under the rule of Theorem~\ref{thm:terminate}, $s\le2^{m(r-1)}$, so the fitted search makes at most
\[
1+16\cdot2^{m(r-1)}
\]
queries. This is finite but exponential; the $O(m\,D\,s)$ bound is polynomial only when $s$ is.
\end{corollary}

\noindent The theorem describes the exact algorithm, not our implementations. Neither keeps the current value on a tie: the NumPy reference takes the smallest minimizing value, and the prototype the first of its candidates. The prototype also minimizes numerically, as described above. Both stop after at most $12$ sweeps, so for them termination is enforced by this cap. No run came near the cap: across the $51$ runs of the validation and scaling studies below, $s\le5$.

\subsection{Measured Cost}\label{sec:measured}
Table~\ref{tab:search} sets each step's bound ($B$ kept valuations, $L_\ell$ and $s$ rounds and sweeps, $D$ frequencies per clock) beside its measured cost at $(k,r)=(14,7)$, where all five steps return $\nu=(28,30,19)$ with energy $-12.153846$\,mHa \codecite{hamilton_vqe/08_search_method_comparison}{search}. Each step removes one factor from the bound: $2^{m(r-1)}$ becomes $3^m$, then $2m$, then a constant per clock, and step~5 drops the coarse seed.

\begin{table}[t]
\caption{Energy queries at $(k,r)=(14,7)$, $m=3$, against each step's bound.}
\label{tab:search}
\centering\small
\begin{tabular}{@{}llrr@{}}
\toprule
step & query bound & queries & fewer than step 1\\
\midrule
1 every point & $2^{m(r-1)}$ & $262{,}144$ & $1\times$\\
2 zoom in & $2^{m(r_0-1)}+B\,3^m(r-r_0)$ & $315$ & $832\times$\\
3 one knob & $2^{m(r_0-1)}+2Bm\sum_\ell L_\ell$ & $189$ & $1{,}387\times$\\
4 fit the wave & $O(2^{m(r_0-1)}+BmDs)$ & $128$ & $2{,}048\times$\\
5 live start & $1+16s$ & $38$ & $6{,}899\times$\\
\bottomrule
\end{tabular}
\end{table}

\begin{figure}[t]
\centering
\begin{tikzpicture}
\begin{groupplot}[group style={group size=2 by 1, horizontal sep=1.5cm},
  width=0.42\linewidth, height=4.6cm, grid=major, grid style={gray!20},
  xlabel={clock precision $r$}, xtick={4,5,6,7,8,9,10},
  tick label style={font=\scriptsize}, label style={font=\footnotesize},
  legend style={font=\scriptsize, draw=none, fill=white}, legend cell align=left]
\nextgroupplot[ymode=log, ymin=5, ymax=1e9, title={\footnotesize (a) energy queries $Q$},
  legend pos=north west]
\addplot[thick, dashed, black!60] coordinates {(4,512) (5,4096) (6,32768) (7,262144) (8,2097152) (9,16777216) (10,134217728)};
\addlegendentry{grid size $2^{3(r-1)}$}
\addplot[thick, black!35, mark=square*, mark size=1.6pt] coordinates {(4,18) (5,29) (6,30) (7,36) (8,38) (9,68) (10,54)};
\addlegendentry{$Q$, $k=8$}
\addplot[thick, black!60, mark=triangle*, mark size=1.6pt] coordinates {(4,18) (5,29) (6,30) (7,38) (8,47) (9,53) (10,53)};
\addlegendentry{$Q$, $k=10$}
\addplot[thick, violet, mark=*, mark size=1.6pt] coordinates {(4,18) (5,29) (6,30) (7,38) (8,47) (9,53) (10,53)};
\addlegendentry{$Q$, $k=14$}
\nextgroupplot[ymin=0, ymax=6, ytick={0,...,6}, title={\footnotesize (b) coordinate sweeps $s$},
  legend pos=south east]
\addplot[thick, black!35, mark=square*, mark size=1.6pt] coordinates {(4,2) (5,3) (6,3) (7,3) (8,3) (9,5) (10,4)};
\addlegendentry{$k=8$}
\addplot[thick, black!60, mark=triangle*, mark size=1.6pt] coordinates {(4,2) (5,3) (6,3) (7,3) (8,4) (9,4) (10,4)};
\addlegendentry{$k=10$}
\addplot[thick, violet, mark=*, mark size=1.6pt] coordinates {(4,2) (5,3) (6,3) (7,3) (8,4) (9,4) (10,4)};
\addlegendentry{$k=14$}
\end{groupplot}
\end{tikzpicture}
\caption{Step~5 at $r=4,\dots,10$ for three $k$: (a) energy queries against grid size; (b) sweeps $s$.}
\Description{Two line plots against clock precision r from 4 to 10. Left, on a log scale: the grid size rises from 512 to 134 million while the query counts for k=8, 10 and 14 stay between 18 and 68. Right: the number of sweeps stays between 2 and 5.}
\label{fig:scaling}
\end{figure}
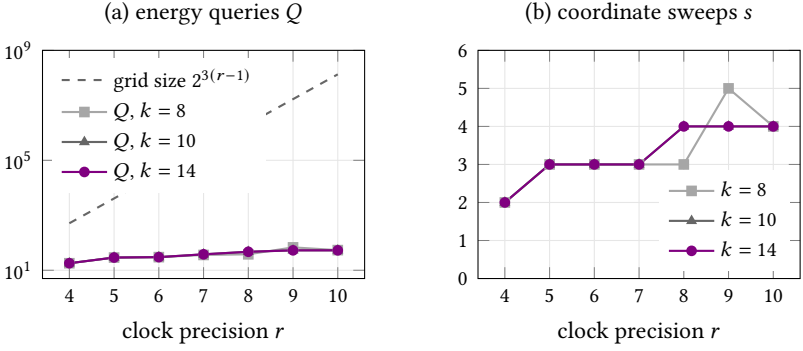

\textbf{Precision.} Fig.~\ref{fig:scaling} runs step~5 for $k\in\{8,10,14\}$ and $r=4,\dots,10$ \codecite{hamilton_vqe/12_analytic_direct_precision_scaling}{scaling}. The grid grows by $2^3$ per bit, from $512$ to $134{,}217{,}728$ valuations, while the queries stay between $18$ and $68$, always within $1+16s$. For $k=14$ the pairs $(s,Q)$ are $(2,18)$, $(3,29)$, $(3,30)$, $(3,38)$, $(4,47)$, $(4,53)$, $(4,53)$. The counts are not monotone in $r$ ($k=8$ needs $68$ queries at $r=9$ and $54$ at $r=10$), because the number of sweeps varies; the bound depends on $s$, not on $r$.

At $(14,10)$ the search starts at $\nu_0=(128,128,128)$ and stops after four sweeps at $(223,238,155)$, the valuation that the constructive bound of Sec.~\ref{sec:errors} predicted, with energy $-12.163132$\,mHa, $0.000119$\,mHa above $E_0(H_{14})$.

The constructive bound only bounds the grid optimum from above,
\[
E_0(H_{14})\;\le\; E^\star_{14,10}\;\le\; E_{H_{14}}(223,238,155),
\]
so we also evaluated all $134{,}217{,}728$ valuations in double precision \codecite{hamilton_vqe/14_exhaustive_r10_certificate}{exhaustive}. $(223,238,155)$ is the grid optimum, and the runner-up, $(223,239,155)$, lies $5.0\times10^{-5}$\,mHa, or $9.7\times10^{-7}$ in scaled units, above it. Each energy is evaluated as a quadratic form in four amplitudes computed from sines and cosines.

This separation is many orders of magnitude larger than the expected double-precision rounding error. Two checks support this. Re-evaluating the three lowest valuations with \texttt{mpmath} at 50 significant digits changes their energies by at most $6\times10^{-17}$ in scaled units and gives the same order and margin. On $200{,}000$ random valuations (seed $0$), double precision and NumPy's \texttt{longdouble}, which is 80-bit x87 extended precision with a 64-bit significand on our x86-64 machine, differ by at most $7.3\times10^{-16}$, again in scaled units. The second is a sample, not a bound over the grid. This is an exhaustive numerical check, not an exact-arithmetic or formally certified proof of grid optimality.

The noncancelling budget is $0.000559$\,mHa, within the $0.001$\,mHa target of Table~\ref{tab:precision}.

\textbf{Reliability.} Exact fits do not make coordinate search a global optimizer. Table~\ref{tab:validation} compares step~5, run by the NumPy reference, with exhaustive search for $k=3,\dots,12$ and $r=4,5,6$, 30 settings in all \codecite{hamilton_vqe/11_analytic_direct_global_validation}{validation}; $r$ starts at $4$ because the sample at $\phi=7\pi/8$ needs it. Step~5 returns the exhaustive optimum in 21 settings, all ten at $r=5$ and nine of ten at $r=6$, but only two of ten at $r=4$. Its last column is the largest energy of the step~5 result above the grid optimum, over the ten values of $k$.

At $r=4$, for $k=4$ and $k=6,\dots,12$, the search stops at $(4,0,3)$ while the optimum is $(3,4,2)$; at $(k,r)=(3,6)$ it stops at $(12,15,6)$ instead of $(11,15,5)$. To confirm that these misses are coordinate traps rather than errors of fitting or minimization, we enumerated every grid value of each clock at each of the nine stopping points, with the other clocks fixed and no fitted curve involved. None lowers the energy, so each stopping point is a coordinatewise minimum, as in Theorem~\ref{thm:terminate}. At $(3,6)$ every single-clock change raises the energy, by at least $0.025$\,mHa.

At $(4,0,3)$ the trap is weaker. There $\phi_1=\pi/2$, so $YI(\alpha_1)$ has turned the first qubit fully to $|1\rangle$, which empties the $|0\rangle$ branch. The energy then does not depend on $\alpha_2$ at all, while changing $\alpha_1$ or $\alpha_3$ alone raises it. The search has emptied a branch that the balanced start populated, and reaching $(3,4,2)$ requires moving $\alpha_1$ and $\alpha_2$ together.

\textbf{Ties.} Some valuations prepare the same state, so their energies are equal exactly, not just numerically. When $\phi_1$ is $0$ or $\pi/2$, one branch of the first qubit is empty and the clock behind it has no effect. On the coarse grid at $r=3$, where $\nu(\alpha_1)\in\{0,1,2,3\}$, half the values of $\alpha_1$ do this. Which of the tied valuations a search reports, or keeps in its beam, then depends on the last bits of the computed energies.

Rerunning the compiler-backed searches on another machine reproduced every energy, and every final valuation, of steps~1, 4 and~5 at $(14,7)$ and of step~5 at $(10,5)$ and $(14,10)$. Steps~2 and~3, which keep coarse valuations in a beam, broke a tie at $r=3$ differently. They reached the same $(28,30,19)$ but needed $325$ and $198$ queries instead of $315$ and $189$; the bound of Proposition~\ref{prop:beam}, $388$, holds for both runs of step~2. The first effect is a property of the program, valuations that prepare the same state. The second is a property of the search, which representative of a tied set it keeps.

\begin{table}[t]
\caption{Step~5 against exhaustive search over $k=3,\dots,12$ (mHa).}
\label{tab:validation}
\centering\small
\begin{tabular}{@{}rrrrr@{}}
\toprule
$r$ & optimum found & mean queries & mean sweeps & largest excess\\
\midrule
4 & $2/10$  & $18.2$ & $2.0$ & $0.476405$\\
5 & $10/10$ & $26.8$ & $2.7$ & $<10^{-12}$\\
6 & $9/10$  & $30.3$ & $3.0$ & $0.050511$\\
\bottomrule
\end{tabular}
\end{table}

\subsection{A Noncommuting Pair and a Free Clock}\label{sec:pair}
The second ansatz (Fig.~\ref{fig:stairs}b) gives each clock its own Pauli generator, $YI(\alpha_1)\parallel ZY(\alpha_2)\parallel IY(\alpha_3)$, a single Hamiltonian process whose grid minimum at $r=7$ occurs at $\nu=(28,3,19)$. $YI$ and $ZY$ anticommute, so our compiler compiles this pair at a Trotter level $n$, as $2^n$ repetitions with each clock divided by $2^n$; $IY$ commutes with both, so it is compiled once, exactly. The syntax shows which clocks need this.

At $n=0$ the circuit runs $YI$, then $ZY$, then $IY$, and by the fusion of Sec.~\ref{sec:fusion} it prepares the state of the first ansatz with controlled clock values $t_3+t_2$ and $t_3-t_2$, where $t_j=\nu(\alpha_j)/2^r$; as $n$ grows it approaches instead the simultaneous motion of $YI\parallel ZY$, a different family, whose optimum is $(28,3,19)$.

The level changes the answer: the optimum of the $n=0$ circuit, $(36,37,57)$, is $33$\,mHa off as a valuation of the program \codecite{trotter_pair}{pair}. Steps 2--3 recover each level's exhaustive optimum for $n\geq1$, though both miss it at $n=0$; each query's circuit grows as $2\cdot2^n+1$ Pauli-evolution factors (atoms).

Step 4 still reads its frequencies off the program: the free clock keeps its three samples at every level, while each clock of the pair repeats $2^n$ times and needs $2^{n+1}+1$, more than the $64$-point axis holds at $n=5$; fitting with the $n=0$ budget instead misses the level's optimum at every $n\geq1$ but one.

The level is also a dimension to search. Valuations do not depend on $n$, so the optimum at level $n-1$ seeds level $n$, as $q\mapsto 2q$ seeds precision $r+1$: step 5 at $n=0$, then one-knob moves, reaches $(28,3,19)$ at $n=4$ for $1{,}709$ atom applications (atoms executed over all queries), against $5{,}379$ for step 3 started cold at $n=4$ (Fig.~\ref{fig:stairs}d, right). One query at $n+1$ suggests when to stop: the energy moves by $26$\,mHa at $n=0$ and by at most $0.007$\,mHa from $n=2$. This is a heuristic, not a certificate; as shown below, even a small error can reorder two nearby valuations.

Running the pair's schedule alternately forwards and backwards \cite{strang1968} (Fig.~\ref{fig:stairs}d, left), with the same atoms, makes the error near the optimum fall about $4\times$ per level instead of $2\times$, and the optimum appears at $n=2$ instead of $n=4$, for $867$ atom applications (Fig.~\ref{fig:stairs}d, middle: solid lines show the error near the optimum, dashed lines the regret, i.e.\ the program's energy at the level's optimum above its own optimum).

A level finds the right valuation once its regret reaches $0$ (stars), and that requires the error not to be small but to be nearly flat: the optimum leads its neighbour $(28,3,20)$ by only $0.018$\,mHa, so at canonical $n=3$ an error of $+0.014$\,mHa on the one and $-0.016$\,mHa on the other already swaps them.

\subsection{Limitations}\label{sec:limits}
The query bounds are proved, but they are conditional on the number of sweeps, which is measured; the evidence comes from a single two-qubit benchmark. Coordinate search is not globally reliable: step~5 misses the grid optimum in 9 of the 30 settings of Table~\ref{tab:validation}, 8 of them at the coarsest precision $r=4$. Our prototype simulates statevectors with $2^N$ amplitudes for $N$ qubits; efficient classical methods, such as stabilizer \cite{aaronson2004} and bounded-width tensor-network simulation \cite{markov2008}, cover only restricted programs.

An energy query is also not a unit of hardware cost. On hardware each query estimates $\langle H_k\rangle$ by measuring $g$ groups of commuting Pauli strings with $S$ shots each, about $QgS$ circuit runs in all \cite{wecker2015}, and our statevector simulation has no shot noise. With noisy energies, the fits of Corollary~\ref{cor:samples} amplify the noise by a factor fixed by the sample design:

\begin{proposition}[Noise in a fit]\label{prop:conditioning}
Let $V\mathbf c=\mathbf y$ be a fit, with $V$ nonsingular, $\mathbf y$ the sampled energies and $\mathbf c$ the coefficients. If the samples are perturbed by $\delta\mathbf y$, the coefficients move by $\delta\mathbf c=V^{-1}\delta\mathbf y$, so $\|\delta\mathbf c\|\le\|V^{-1}\|\,\|\delta\mathbf y\|$ and $\|\delta\mathbf c\|/\|\mathbf c\|\le\kappa(V)\,\|\delta\mathbf y\|/\|\mathbf y\|$, with $\kappa(V)=\|V\|\|V^{-1}\|$.
\end{proposition}

\noindent For our sample designs $\kappa(V)=1+\sqrt2\approx2.414$ for $YI(\alpha_1)$ and $\kappa(V)\approx28.30$ for the controlled clocks \codecite{hamilton_vqe/09_harmonic_fit_conditioning}{conditioning}. With exact energies this is harmless; with sampled ones, more samples fitted by least squares, or samples spread more evenly, would reduce the second.

\section{Conclusion}\label{sec:conclusion}
\VH adds one construct to \Ham, the open clock, filled by a valuation. Its programs run: each valuation compiles through \Ham to a circuit. And we can prove things about them: because the semantics stays exact and affine in the clocks, Hamiltonian equivalence for every valuation is a finite calculation, which Agda checks for a fragment, in a Pauli-sum semantics whose rules we checked numerically against the matrices.

The same exactness carries over to a chemistry eigensolver: its error splits exactly between the precision of the target and that of the clocks, each with an a priori bound, and the program fixes the shape of each clock's energy curve, so a search with a constant number of queries per clock reaches the optimum of a grid of 134 million valuations, $0.000119$\,mHa above the ground energy of the rounded target, in 53 queries, as exhaustive numerical evaluation confirms, though, as a coordinate search, it is not a global optimizer.

The evidence is of three kinds and should be read as such: the equivalences, including block fusion as a parallel composition, are checked by Agda; the step from parallel to sequence, and the error and query bounds, are pen-and-paper proofs; and the prototype compiler and the measured costs are tested on one benchmark, not verified.

Future work is to extend \Ham's sound and complete calculus to all of \VH with respect to Hamiltonian equivalence, mechanize it, and verify compiler rewrites.

\bibliographystyle{ACM-Reference-Format-citeorder}
\bibliography{references}

\raggedbottom
\appendix
\clearpage
\section{The Agda Development}\label{app:agda}
The core of \texttt{VHamilton.agda} \codeagda, with Unicode rendered as mathematics. An open clock $\pm\delta_k^\alpha$ is written \texttt{\(\pm\delta\)\^{}\,\(\alpha\)}, with each variable's $k$ supplied alongside the valuation. The identity Pauli is written $\mathbf{1}$. A Hamiltonian is a list of Pauli strings with affine weights \texttt{LF m}, a rational constant plus one rational coefficient per variable.

\noindent\begin{codebox}
data Pauli : Set where \(\mathbf{1}\) X Y Z : Pauli
PString = Vec Pauli
LF m    = \(\mathbb{Q}\times\) Vec \(\mathbb{Q}\) m    -- affine form in the clock values
Ham m n = List (LF m \(\times\) PString n)

-- d\(\sb{0}\) \(\oplus\) d\(\sb{1}\) = \(\mathbf{1}\) \(\otimes\) (d\(\sb{0}\)+d\(\sb{1}\))/2 + Z \(\otimes\) (d\(\sb{0}\)-d\(\sb{1}\))/2
split (c\(\sb{0}\) , P) (c\(\sb{1}\) , _) = (half (c\(\sb{0}\) \(\boxplus\) c\(\sb{1}\)) , \(\mathbf{1}\) \(\dblcolon\) P)
                    \(\dblcolon\) (half (c\(\sb{0}\) \(\boxplus\) neg c\(\sb{1}\)) , Z \(\dblcolon\) P) \(\dblcolon\) []
diag (leaf s)  = (valS s , []) \(\dblcolon\) []
diag (d\(\sb{0}\) \(\oplus\) d\(\sb{1}\)) = concat (zipWith split (diag d\(\sb{0}\)) (diag d\(\sb{1}\)))

conjT u (c , P) with cj u P    -- conjugation table: u P u\(\sp\dagger\) = \(\pm\)P'
... | true  , P' = c , P'
... | false , P' = neg c , P'

\(\llbracket\) bare d   \(\rrbracket\)A = diag d
\(\llbracket\) conj u a \(\rrbracket\)A = map (conjT u) \(\llbracket\) a \(\rrbracket\)A
\(\llbracket\) act a \(\rrbracket\)   = \(\llbracket\) a \(\rrbracket\)A
\(\llbracket\) p \(\parallel\) q \(\rrbracket\) = \(\llbracket\) p \(\rrbracket\) ++ \(\llbracket\) q \(\rrbracket\)

coef h Q = foldr step lf0 h    -- total weight on the string Q
  where step (c , P) acc = if eqS P Q then c \(\boxplus\) acc else acc

p \(\simeq\) q = \(\forall\) Q \(\to\) coef \(\llbracket\) p \(\rrbracket\) Q \(\equiv\) coef \(\llbracket\) q \(\rrbracket\) Q
sound : p \(\simeq\) q \(\to\) \(\forall\) k \(\nu\) Q \(\to\) \(\llbracket\) p \(\rrbracket\)^ k \(\nu\) Q \(\equiv\) \(\llbracket\) q \(\rrbracket\)^ k \(\nu\) Q
sound eq k \(\nu\) Q = cong (evalLF k \(\nu\)) (eq Q)

ZZ       = act (bare ((leaf (\(\delta\)^ \(\gamma\)) \(\oplus\) leaf (-\(\delta\)^ \(\gamma\)))
                    \(\oplus\) (leaf (-\(\delta\)^ \(\gamma\)) \(\oplus\) leaf (\(\delta\)^ \(\gamma\)))))
IZ       = bare ((leaf (\(\delta\)^ \(\gamma\)) \(\oplus\) leaf (-\(\delta\)^ \(\gamma\)))
                \(\oplus\) (leaf (\(\delta\)^ \(\gamma\)) \(\oplus\) leaf (-\(\delta\)^ \(\gamma\))))
CX-IZ-CX = act (conj CX IZ)

ZZ\(\simeq\)CX-IZ-CX : ZZ \(\simeq\) CX-IZ-CX
ZZ\(\simeq\)CX-IZ-CX (\(\mathbf{1}\) \(\dblcolon\) _ \(\dblcolon\) []) = refl
ZZ\(\simeq\)CX-IZ-CX (X \(\dblcolon\) _ \(\dblcolon\) []) = refl
ZZ\(\simeq\)CX-IZ-CX (Y \(\dblcolon\) _ \(\dblcolon\) []) = refl
ZZ\(\simeq\)CX-IZ-CX (Z \(\dblcolon\) _ \(\dblcolon\) []) = refl
\end{codebox}

\clearpage
\section{Proofs}\label{app:proofs}
Throughout, $\|\cdot\|$ is the operator norm, and states are normalized.

\begin{proof}[Proof of Proposition~\ref{prop:identity}]
$H$ and $c\id$ commute, so
\[
e^{-it(H+c\id)}=e^{-itH}e^{-itc\id}=e^{-itc}e^{-itH}.
\]
For the energy,
\[
\langle\psi|H+c\id|\psi\rangle=\langle\psi|H|\psi\rangle+c\langle\psi|\psi\rangle=\langle\psi|H|\psi\rangle+c.
\]
\end{proof}

\begin{proof}[Proof of Proposition~\ref{prop:affine}]
By linearity, $\langle\psi|aH+b\id|\psi\rangle=a\langle\psi|H|\psi\rangle+b$ for every normalized $|\psi\rangle$, in particular for $|\psi(\nu)\rangle$. The map $x\mapsto ax+b$ with $a>0$ is strictly increasing, so it preserves the order of the energies of any two valuations, hence the set of minimizers. Applying the same map to all normalized states,
\[
E_0(\widetilde H)=\min_{\psi}\big(a\langle\psi|H|\psi\rangle+b\big)=aE_0(H)+b.
\]
\end{proof}

\begin{proof}[Proof of Lemma~\ref{lem:open}]
By induction on $p$; the induction passes through bare actions and actions, and $\nu$ is fixed throughout. Write $t=\nu(\alpha)/2^r$. For a clock,
\[
\llbracket\pm\delta_{k+r}^\alpha\rrbracket_H^\nu=\pm\frac{\nu(\alpha)}{2^{k+r}}=t\cdot(\pm2^{-k})=t\,\llbracket\pm\delta_k\rrbracket_H ,
\]
and $0$ is unchanged, with $0=t\cdot0$. Opening commutes with every other construct, $(d_1\oplus d_2)^\alpha_r=(d_1)^\alpha_r\oplus(d_2)^\alpha_r$ and so on, and leaves transformations untouched. So for a direct sum, with $D_i=\llbracket d_i\rrbracket_H$ and the induction hypotheses $\llbracket(d_i)^\alpha_r\rrbracket_H^\nu=tD_i$, we get
\[
\llbracket(d_1\oplus d_2)^\alpha_r\rrbracket_H^\nu=tD_1\oplus tD_2=t(D_1\oplus D_2)=t\llbracket d_1\oplus d_2\rrbracket_H .
\] For a conjugation and for parallel composition,
\[
\llbracket u\rrbracket_U(tA)\llbracket u\rrbracket_U^\dagger=t\,\llbracket u\rrbracket_UA\llbracket u\rrbracket_U^\dagger,\qquad tP+tQ=t(P+Q),
\]
with $A,P,Q$ the Hamiltonians of the unopened parts.
\end{proof}

\begin{proof}[Proof of Lemma~\ref{lem:closed}]
A Hamiltonian $G$ with $G^2=\Pi$ a projector (including $\Pi=\id$) satisfies
\[
e^{-i\phi G}=(\id-\Pi)+\cos\phi\,\Pi-i\sin\phi\,G,
\]
as both sides agree on the range of $\Pi$, where $G$ acts as an involution, and on its kernel, where $G=0$.

With $\llbracket YI(\alpha_1)\rrbracket_U^\nu=e^{-i\phi_1\,Y\otimes\id}$ and $Y|0\rangle=i|1\rangle$, the first block gives $\cos\phi_1|00\rangle+\sin\phi_1|10\rangle$. The second, $e^{-i\phi_2\,|0\rangle\langle0|\otimes Y}$, acts as $e^{-i\phi_2Y}$ on the second qubit of the $|0\rangle$ branch and as the identity on the $|1\rangle$ branch, taking $|00\rangle$ to $\cos\phi_2|00\rangle+\sin\phi_2|01\rangle$. The third does the same on the $|1\rangle$ branch with $\phi_3$, taking $|10\rangle$ to $\cos\phi_3|10\rangle+\sin\phi_3|11\rangle$. Collecting amplitudes gives the stated vector.
\end{proof}

\begin{proof}[Proof of Theorem~\ref{thm:express}]
Let $v=(a,b,c,d)$ be a real unit vector, and let $\rho_0=\sqrt{a^2+b^2}$ and $\rho_1=\sqrt{c^2+d^2}$, so $\rho_0^2+\rho_1^2=1$.

Replacing $v$ by $-v$ if necessary, we may assume that $(c,d)=\rho_1(\cos\phi_3,\sin\phi_3)$ for some $\phi_3\in[0,\pi)$: a nonzero $(c,d)$ has a polar angle $\theta\in[0,2\pi)$, and if $\theta\ge\pi$ then $-(c,d)$ has angle $\theta-\pi\in[0,\pi)$; if $(c,d)=0$, take $\phi_3=0$.

With this sign fixed, the same argument applied to $(a,b)$, now recording the sign instead of changing $v$, gives $(a,b)=\sigma\rho_0(\cos\phi_2,\sin\phi_2)$ with $\phi_2\in[0,\pi)$ and $\sigma\in\{1,-1\}$; if $(a,b)=0$, take $\phi_2=0$ and $\sigma=1$.

Finally let $\phi_1=\arcsin\rho_1\in[0,\pi/2]$ if $\sigma=1$ and $\phi_1=\pi-\arcsin\rho_1$ if $\sigma=-1$. In both cases $\sin\phi_1=\rho_1$ and $\cos\phi_1=\sigma\rho_0$, and $\phi_1\in[0,\pi)$, unless $\sigma=-1$ and $\rho_1=0$, which would give $\phi_1=\pi$. In that case $v=(a,b,0,0)$, so $-v$ is fixed by the first step just as well, and taking it instead turns $\sigma$ to $1$. Substituting into Lemma~\ref{lem:closed} gives
\[
(\sigma\rho_0\cos\phi_2,\ \sigma\rho_0\sin\phi_2,\ \rho_1\cos\phi_3,\ \rho_1\sin\phi_3)=(a,b,c,d).
\]
\end{proof}

\begin{proof}[Proof of Proposition~\ref{prop:seqpar}]
Write $P=\llbracket p\rrbracket_H^\nu$ and $Q=\llbracket q\rrbracket_H^\nu$. For commuting matrices, $e^{A}e^{B}=e^{A+B}$, since the binomial expansion of $(A+B)^m$ is then valid term by term in the exponential series. Hence
\[
\llbracket q\rrbracket_U^\nu\llbracket p\rrbracket_U^\nu=e^{-2\pi iQ}e^{-2\pi iP}=e^{-2\pi i(P+Q)}=e^{-2\pi i\llbracket p\parallel q\rrbracket_H^\nu}.
\]
\end{proof}

\begin{proof}[Proof of Corollary~\ref{cor:fusion}]
The Hamiltonians are $t_2\,|0\rangle\langle0|\otimes Y$ and $t_3\,|1\rangle\langle1|\otimes Y$, with $t_j=\nu(\alpha_j)/2^r$. Their products in either order contain the factor $|0\rangle\langle0|1\rangle\langle1|=0$ or $|1\rangle\langle1|0\rangle\langle0|=0$, so they commute. By Proposition~\ref{prop:seqpar} the sequence equals $\llbracket CY(\alpha_2)\parallel\overline CY(\alpha_3)\rrbracket_U^\nu$. This in turn equals $\llbracket\mathit{Mux}(\alpha_2,\alpha_3)\rrbracket_U^\nu$, because Hamiltonian equivalence gives equal Hamiltonians for every valuation, hence equal exponentials. The ansatz statement follows by applying both sides to $\llbracket YI(\alpha_1)\rrbracket_U^\nu|00\rangle$.
\end{proof}

\begin{proof}[Proof of Proposition~\ref{prop:decomp}]
The identity is
\[
E^\star_{k,r}-E_0(\bar H)=\big(E^\star_{k,r}-E_0(H_k)\big)+\big(E_0(H_k)-E_0(\bar H)\big).
\] By the variational principle, every normalized state has $\langle\psi|H_k|\psi\rangle\ge E_0(H_k)$, the least eigenvalue, so in particular $E^\star_{k,r}\ge E_0(H_k)$ and $\delta_r\ge0$. The inequality is the triangle inequality, with $|\delta_r|=\delta_r$.
\end{proof}

\begin{proof}[Proof of Theorem~\ref{thm:hround}]
Write $H_k=\bar H+\Delta$ with $\Delta=\sum_\ell(c^{(k)}_\ell-\bar c_\ell)P_\ell$. By the variational characterization, $E_0(H_k)=\min_\psi\langle\psi|\bar H+\Delta|\psi\rangle$, and $|\langle\psi|\Delta|\psi\rangle|\le\|\Delta\|$ for every normalized $\psi$. So
\[
E_0(\bar H)-\|\Delta\|\;\le\; E_0(H_k)\;\le\; E_0(\bar H)+\|\Delta\|,
\]
which is Weyl's inequality for the least eigenvalue. Every Pauli string is unitary and Hermitian, so $\|P_\ell\|=1$, and the triangle inequality gives $\|\Delta\|\le\sum_\ell|c^{(k)}_\ell-\bar c_\ell|$. Rounding to the nearest multiple of $2^{-k}$ moves each coefficient by at most $2^{-(k+1)}$, so $\|\Delta\|\le M2^{-(k+1)}$.
\end{proof}

\begin{proof}[Proof of Lemma~\ref{lem:closedgrid}]
Write $\psi(\phi_1,\phi_2,\phi_3)$ for the state of Lemma~\ref{lem:closed}. A global sign does not change the energy, and three symmetries of the closed form follow from $\cos(\phi+\pi)=-\cos\phi$, $\sin(\phi+\pi)=-\sin\phi$, $\cos(\pi-\phi)=-\cos\phi$ and $\sin(\pi-\phi)=\sin\phi$:
\begin{align*}
&\text{(i)} & \psi(\phi_1+\pi,\phi_2,\phi_3)&=-\psi(\phi_1,\phi_2,\phi_3),\\
&\text{(ii)} & \psi(\phi_1,\pi,\phi_3)&=(-\cos\phi_1,\,0,\,\sin\phi_1\cos\phi_3,\,\sin\phi_1\sin\phi_3)=\psi(\pi-\phi_1,0,\phi_3),\\
&\text{(iii)} & \psi(\phi_1,\phi_2,\pi)&=(\cos\phi_1\cos\phi_2,\,\cos\phi_1\sin\phi_2,\,-\sin\phi_1,\,0)=\psi(-\phi_1,\phi_2,0).
\end{align*}
In clock values, $t_j=\phi_j/2\pi$, a point of the closed grid has each $t_j=q_j/2^r$ with $0\le q_j\le2^{r-1}$. If $q_2=2^{r-1}$, (ii) replaces it by $q_2=0$ and $q_1$ by $2^{r-1}-q_1$. If $q_3=2^{r-1}$, (iii) replaces it by $q_3=0$ and $q_1$ by $-q_1$. Finally (i) reduces $q_1$ modulo $2^{r-1}$. Each step keeps the energy and every coordinate a multiple of $2^{-r}$, and the result lies in $\{0,\dots,2^{r-1}-1\}^3$.
\end{proof}

\begin{proof}[Proof of Theorem~\ref{thm:rround}]
Write the state and its energy as
\[
|\psi(t)\rangle=V_3V_2V_1|00\rangle,\quad V_j=e^{-2\pi it_jG_j},\qquad E(t)=\langle\psi(t)|H_k|\psi(t)\rangle,
\]
a smooth function on $\mathbb R^3$.

\emph{Derivatives.} Fix $j$ and let
\[
|\eta_j\rangle=V_j\cdots V_1|00\rangle,\qquad A_j=V_{j+1}^\dagger\cdots V_3^\dagger H_kV_3\cdots V_{j+1},
\]
so that $A_j$ does not depend on $t_1,\dots,t_j$ and $\|A_j\|=\|H_k\|$. Then $E=\langle\eta_j|A_j|\eta_j\rangle$, and differentiating $V_j$ gives
\[
\partial_jE=2\pi\langle\eta_j|B_j|\eta_j\rangle
\]
with $B_j=i[G_j,A_j]$ Hermitian and $\|B_j\|\le2\|G_j\|\|A_j\|=2\|H_k\|$. So $|\partial_jE|\le4\pi\|H_k\|$. For $l\le j$, the factor $B_j$ does not depend on $t_l$, and $\langle\eta_j|B_j|\eta_j\rangle=\langle\eta_l|W^\dagger B_jW|\eta_l\rangle$ with $W=V_j\cdots V_{l+1}$ independent of $t_l$. The same computation gives $|\partial_l\partial_jE|\le2\pi\cdot2\pi\cdot2\|B_j\|\le16\pi^2\|H_k\|$; by symmetry of second derivatives this covers $l>j$ too. Hence the Hessian satisfies, everywhere,
\[
\|\nabla^2E\|\le\|\nabla^2E\|_F\le16\pi^2m\|H_k\|.
\]

\emph{A continuous optimum.} By Theorem~\ref{thm:express} the real ground state of $H_k$ is $\pm\psi(t^\ast)$ for some $t^\ast\in[0,\frac12)^3$, so $E(t^\ast)=E_0(H_k)$, and since $E\ge E_0(H_k)$ everywhere, $t^\ast$ is a global minimum of $E$ on $\mathbb R^3$, with $\nabla E(t^\ast)=0$. Round each $t^\ast_j$ to the nearest multiple of $2^{-r}$, giving $\tilde t\in[0,\frac12]^3$ with $|\tilde t_j-t^\ast_j|\le2^{-(r+1)}$. By Lemma~\ref{lem:closedgrid}, $E(\tilde t)$ is the energy of some grid valuation, so $\delta_r(k,r)\le E(\tilde t)-E(t^\ast)$.

\emph{First order.} Moving from $t^\ast$ to $\tilde t$ one coordinate at a time, the mean value theorem gives
\[
E(\tilde t)-E(t^\ast)\le\sum_j4\pi\|H_k\|\,2^{-(r+1)}=2\pi m\|H_k\|2^{-r}.
\]

\emph{Second order.} Taylor's theorem with the Lagrange remainder, and $\nabla E(t^\ast)=0$, give, for some $\xi$ on the segment,
\[
E(\tilde t)-E(t^\ast)=\tfrac12(\tilde t-t^\ast)^{\mathsf T}\nabla^2E(\xi)(\tilde t-t^\ast)\le\tfrac12\cdot16\pi^2m\|H_k\|\cdot m\,2^{-2(r+1)}=2\pi^2m^2\|H_k\|4^{-r}.
\]
\end{proof}

\begin{proof}[Proof of Theorem~\ref{thm:nested}]
If $\nu(\alpha_j)=q_j$ is on the grid at precision $r$, then $2q_j\le2^r-2$ is on the grid at precision $r+1$ and gives the same clock value, $q_j/2^r=2q_j/2^{r+1}$, hence the same energy. The minimum over the larger grid is therefore no larger, and subtracting $E_0(H_k)$ gives $\delta_r(k,r+1)\le\delta_r(k,r)$. The limit follows from $0\le\delta_r(k,r)\le2\pi^2m^2\|H_k\|4^{-r}$.
\end{proof}

\begin{proof}[Proof of Proposition~\ref{prop:construct}]
$E^\star_{k,r}\le E_{H_k}(\tilde\nu)$ because $\tilde\nu$ is on the grid, and $E^\star_{k,r}\ge E_0(H_k)$ by the variational principle. Subtract $E_0(H_k)$.
\end{proof}

\begin{proof}[Proof of Proposition~\ref{prop:beam}]
The seed evaluates all $2^{m(r_0-1)}$ valuations at precision $r_0$. Doubling a valuation keeps its clock values and so its energy. At each of the $r-r_0$ finer precisions, each of the $B$ kept valuations has $3^m$ valuations within one step in every clock, itself included, so at most $B3^m$ are evaluated; caching repeated ones can only lower the count.
\end{proof}

\begin{proof}[Proof of Proposition~\ref{prop:coord}]
As for Proposition~\ref{prop:beam}, except that a round from one valuation evaluates only the at most $2m$ valuations that change one clock by one step.
\end{proof}

\begin{proof}[Proof of Theorem~\ref{thm:spectral}]
Write the ansatz state as $W\,e^{-i\phi G}|\eta\rangle$, with $|\eta\rangle$ produced by the earlier blocks and $W$ the later ones, which do not depend on $\phi$; so
\[
f(\phi)=\langle\eta|e^{i\phi G}Ae^{-i\phi G}|\eta\rangle,\qquad A=W^\dagger H_kW.
\] Substituting $e^{-i\phi G}=\sum_\mu e^{-i\mu\phi}\Pi_\mu$ and its adjoint gives the double sum. Group its terms by $\omega=\lambda-\mu$. The terms with $\omega=0$ give a real constant $c$, since $f$ and the other groups are real. Since $A$ is Hermitian, the coefficient $\langle\eta|\Pi_\mu A\Pi_\lambda|\eta\rangle$ of $e^{-i\omega\phi}$ is the complex conjugate of the coefficient $\langle\eta|\Pi_\lambda A\Pi_\mu|\eta\rangle$ of $e^{i\omega\phi}$. So for each $\omega>0$ the two groups add to
\[
2\operatorname{Re}(z_\omega e^{i\omega\phi})=a_\omega\cos\omega\phi+b_\omega\sin\omega\phi,
\]
with $z_\omega$ the summed coefficient, $a_\omega=2\operatorname{Re}z_\omega$ and $b_\omega=-2\operatorname{Im}z_\omega$.
\end{proof}

\begin{proof}[Proof of Corollary~\ref{cor:samples}]
For $YI(\alpha_1)$, $f(\phi)=c+a_2\cos2\phi+b_2\sin2\phi$, so $f(0)=c+a_2$, $f(\pi/2)=c-a_2$ and $f(\pi/4)=c+b_2$, which give the stated formulas. For a controlled clock,
\[
f(\phi)=c+a_1\cos\phi+b_1\sin\phi+a_2\cos2\phi+b_2\sin2\phi.
\]
Then $e^{2i\phi}f(\phi)=P(e^{i\phi})$ for a complex polynomial $P$ of degree at most $4$, whose coefficients determine $c,a_1,b_1,a_2,b_2$. The five sample phases are distinct in $[0,2\pi)$, so the points $e^{i\phi}$ are distinct, and five values determine $P$. In clock values the samples are $\nu=0$, $2^{r-3}$, $2^{r-2}$, $3\cdot2^{r-3}$ and $7\cdot2^{r-4}$. For $r\ge4$ these are integers below $2^{r-1}$, so they lie on the half-range grid, and none depends on anything but $r$.
\end{proof}

\begin{proof}[Proof of Proposition~\ref{prop:analyticbeam}]
The seed costs $2^{m(r_0-1)}$ queries. Lifting keeps clock values, $q/2^{r_0}=2^{r-r_0}q/2^r$, and so costs none. Each sweep from each of the $B$ lifted valuations fits each of the $m$ clocks from at most $2D+1$ samples by Corollary~\ref{cor:samples}, and queries one chosen valuation.
\end{proof}

\begin{proof}[Proof of Lemma~\ref{lem:seed}]
$\nu_0(\alpha_j)/2^r=1/8$, so $\phi_j=\pi/4$, and every amplitude in Lemma~\ref{lem:closed} is a product of two factors $1/\sqrt2$, that is $1/2$, which gives $|{+}{+}\rangle$. For any state the four probabilities are nonnegative and sum to $1$, so the smallest is at most $1/4$; here all four equal $1/4$.
\end{proof}

\begin{proof}[Proof of Theorem~\ref{thm:direct}]
The start costs one query. In each sweep, clock $j$ is fitted from $2|\Omega_j|+1$ samples (Corollary~\ref{cor:samples}), the fitted curve is minimized without queries, and one query evaluates the chosen valuation, $2|\Omega_j|+2$ in all. With $|\Omega_1|=1$ and $|\Omega_2|=|\Omega_3|=2$ this is $4+6+6=16$ per sweep. Cached repeats only lower the count.
\end{proof}

\begin{proof}[Proof of Theorem~\ref{thm:terminate}]
A step chooses a minimizer along one clock, and the current value is a candidate, so the energy does not increase. Under the tie rule a clock changes only if its current value is not a minimizer, in which case the energy strictly decreases. Valuations at which the search starts a sweep are therefore visited with strictly decreasing energies, except for the last one, and so are pairwise distinct. As the grid is finite, there are finitely many sweeps. The final sweep changes no clock, so at $\nu^\ast$ each clock's current value minimizes the energy along that clock with the others fixed.
\end{proof}

\begin{proof}[Proof of Corollary~\ref{cor:conditional}]
By the proof of Theorem~\ref{thm:terminate}, the $s$ valuations at which sweeps start are distinct points of the grid, which has $2^{m(r-1)}$ points; Theorem~\ref{thm:direct} converts sweeps to queries.
\end{proof}

\begin{proof}[Proof of Proposition~\ref{prop:conditioning}]
Subtract $V\mathbf c=\mathbf y$ from $V(\mathbf c+\delta\mathbf c)=\mathbf y+\delta\mathbf y$ and multiply by $V^{-1}$. For the relative bound, $\|\mathbf y\|\le\|V\|\|\mathbf c\|$. The value $1+\sqrt2$ is the ratio of the largest to the smallest singular value of the $3\times3$ design for $YI(\alpha_1)$, whose rows are $(1,1,0)$, $(1,0,1)$ and $(1,-1,0)$; $V^{\mathsf T}V$ has eigenvalues $2+\sqrt2$, $2$ and $2-\sqrt2$, so
\[
\kappa(V)=\sqrt{\frac{2+\sqrt2}{2-\sqrt2}}=\frac{2+\sqrt2}{\sqrt2}=1+\sqrt2.
\] The value for the controlled clocks is computed numerically.
\end{proof}

\clearpage
\section{Compiler Correctness}\label{app:compiler}
A variational search trusts the compiler: if a clock value were compiled as a turn about the wrong axis, or in the wrong direction, the optimizer would search the wrong landscape and still return an answer. This appendix states the two ordering equations the prototype's compiler follows, shows why tests at unit coefficient cannot check them, and reports the regression suite that does \codecite{hamilton_vqe/02_compiler_regression}{regression}.

\textbf{Order.} Circuits run left to right, while matrices compose right to left:

\begin{lemma}[Chronological order]\label{lem:chrono}
If a circuit applies $V$ and then $U$, it implements $UV$. Hence, since $\llbracket u\cdot v\rrbracket_U=\llbracket u\rrbracket_U\llbracket v\rrbracket_U$, a compiler that lists gates in the order they run must emit
\[
\mathsf{compile}(u\cdot v)=\mathsf{compile}(v);\mathsf{compile}(u).
\]
\end{lemma}

\begin{proposition}[Chronological conjugation]\label{prop:chronoconj}
For a unitary $U$ and Hermitian $A$, $e^{-itUAU^\dagger}=Ue^{-itA}U^\dagger$ for every $t$. Hence
\[\mathsf{compile}(uau^\dagger)=\mathsf{compile}(u^\dagger);\mathsf{compile}(a);\mathsf{compile}(u).\]
\end{proposition}

\begin{proof}
A state $|\psi\rangle$ acted on by $V$ and then $U$ becomes $U(V|\psi\rangle)=(UV)|\psi\rangle$. For the conjugation, $(UAU^\dagger)^j=UA^jU^\dagger$ since $U^\dagger U=\id$, so the exponential series gives $e^{-itUAU^\dagger}=Ue^{-itA}U^\dagger$; read chronologically, this applies $U^\dagger$, then $e^{-itA}$, then $U$.
\end{proof}

\noindent Both equations are what the compiler implements, in the cases for $u\cdot v$ and for conjugation. For a variational program the second matters especially: it decides whether a clock value turns the state along the intended axis, as with $Y=(SH)Z(SH)^\dagger$ in every block of the chemistry ansatz.

\textbf{Why unit coefficients prove nothing.} The obvious tests compile a Pauli evolution and compare it with the exact matrix exponential. At unit coefficient they cannot fail:

\begin{lemma}[Unit Pauli evolutions are degenerate]\label{lem:unitpauli}
For every Pauli string $P$, $e^{-2\pi iP}=\id$. So the unit-time evolutions of $X$, $Y$, $Z$, $-Y$, and of any other Pauli string or its negative, all coincide.
\end{lemma}

\begin{proof}
A Pauli string is Hermitian and squares to $\id$, so its eigenvalues are $\pm1$, and $e^{-2\pi i(\pm1)}=1$ on both eigenspaces. The same holds for $-P$.
\end{proof}

\noindent This is the observation of Sec.~\ref{sec:hamilton} that $Z_0$ makes two full turns. A compiler that turned about the wrong axis, or backwards, would pass every such test.

\textbf{The regression suite.} The suite therefore compiles evolutions at coefficient $1/8$, where axes and signs are visible, and compares them with the exact exponentials up to a global phase:
\begin{itemize}
\item \emph{Compiled evolutions}: $X/8$, $Y/8$, $Z/8$, $XI/8$, $IX/8$, $XY/8$, $YX/8$ and $ZZ/8$, with phase-aligned Frobenius distance at most $1.2\times10^{-15}$, against a tolerance of $10^{-9}$.
\item \emph{Wire order}: the Hamiltonians of the two-qubit labels $XI$, $IX$, $XY$, $YX$ and $ZZ$ equal the Kronecker products with the first Pauli on the first qubit, to within $1.2\times10^{-16}$.
\item \emph{Open clocks}: a program's free variables are found; halving $\pm\delta_1^\alpha$ twice gives $\pm\delta_3^\alpha$; a valuation sets the sign and size of each phase; and the open program $(SH)(\delta_1^\alpha\oplus-\delta_1^\alpha)(SH)^\dagger$ compiles to its exact exponential for a fixed valuation.
\item \emph{End to end}: a one-clock VQE, $Y(\alpha)$ at precision $r=4$ against $X+Z$, whose energy $\sin(\pi\nu(\alpha)/4)+\cos(\pi\nu(\alpha)/4)$ is minimized at $\nu(\alpha)=5$ with energy $-\sqrt2$, the exact ground energy. A $Y$ turn in the wrong direction moves this minimum.
\end{itemize}
All 22 checks pass; we reran the suite for this paper. To check that the suite has teeth, we reversed each equation in turn in the compiler: emitting $u$ before $v$ for $u\cdot v$, or $u$ before $u^\dagger$ for a conjugation. In both cases the $Y/8$ test fails, at distance $\sqrt2$, because $Y=(SH)Z(SH)^\dagger$ is compiled as $X$. By Lemma~\ref{lem:unitpauli}, the same tests at coefficient $1$ would pass both faulty compilers.

\textbf{The Trotter level in the chemistry ansatz.} Sec.~\ref{sec:ansatz} argues that the Trotter level plays no part in the chemistry ansatz. Running the exhaustive search at $(k,r)=(10,3)$ through the compiler at levels $n=0,1,2$ returns the same optimal energy, equal to within $10^{-11}$\,mHa \codecite{hamilton_vqe/04_current_ansatz_trotter_invariance}{levels}. The reported valuation is $(2,0,1)$, except that our rerun reported the tied $(2,2,1)$ at $n=1$: at $\phi_1=\pi/2$ the $|0\rangle$ branch is empty, so all four values of $\alpha_2$ give the same energy (Sec.~\ref{sec:measured}).

\clearpage
\section{Reproducibility}\label{app:repro}
All code and results are in the repository \cite{vhamiltoncode}: the Haskell prototype of \VH (compiler to OpenQASM, search methods), NumPy and Qiskit scripts that serve as independent references, and the Agda development. Each folder under \href{\coderesults}{\texttt{results/}} contains the command that produced it, its raw logs, and machine-readable summaries; every number in this paper is copied from one of them. Table~\ref{tab:repro} maps the paper to the folders, and the commands below regenerate them from the repository root. The prototype needs GHC~9.4 with \texttt{cabal}; the scripts need Python~3 with NumPy, SciPy, Qiskit~2.6 and \texttt{qiskit\_qasm3\_import}; the proofs need Agda~2.6.3 with agda-stdlib~1.7.3. Large exhaustive references are computed by the NumPy scripts rather than by starting the Qiskit oracle once per valuation.

\begin{table}[h]
\caption{Where each result comes from: folders under \texttt{results/} of \cite{vhamiltoncode}.}
\label{tab:repro}
\centering\footnotesize
\newcommand{\rf}[1]{\href{\coderesults #1}{\texttt{#1}}}
\begin{tabular}{@{}>{\raggedright\arraybackslash}p{0.4\linewidth}>{\raggedright\arraybackslash}p{0.56\linewidth}@{}}
\toprule
In the paper & Folder\\
\midrule
Sec.~\ref{sec:equivalence}, Fig.~\ref{fig:agda}, App.~\ref{app:agda} & \href{https://github.com/JavariaGhafoor/vhamilton/tree/main/agda}{\texttt{agda}}\\
QAOA, Fig.~\ref{fig:qaoa} & \rf{qaoa\_one\_edge}\\
Noncommuting block, Table~\ref{tab:trotter} & \rf{trotter\_vqe}\\
Target, Table~\ref{tab:pauli} & \rf{hamilton\_vqe/03\_chemistry\_target\_and\_identity}\\
Trotter level in the ansatz, App.~\ref{app:compiler} & \rf{hamilton\_vqe/04\_current\_ansatz\_trotter\_invariance}\\
Precision selection, Table~\ref{tab:precision} & \rf{hamilton\_vqe/05\_pre\_vqe\_precision\_selection}\\
$(k,r)=(10,3)$ and $(10,5)$, Sec.~\ref{sec:errors} & \rf{hamilton\_vqe/06\_k\_r\_error\_separation}\\
Exhaustive $(k,r)$ sweep, Sec.~\ref{sec:errors} & \rf{hamilton\_vqe/07\_exhaustive\_k\_r\_sweep}\\
Bounds against measurements, Fig.~\ref{fig:errors} & \rf{hamilton\_vqe/13\_error\_calculus\_bounds}\\
Five steps, Fig.~\ref{fig:stairs}c, Table~\ref{tab:search} & \rf{hamilton\_vqe/08\_search\_method\_comparison}\\
Fit conditioning, Sec.~\ref{sec:limits} & \rf{hamilton\_vqe/09\_harmonic\_fit\_conditioning}\\
Step~5 at $(14,10)$, Sec.~\ref{sec:measured} & \rf{hamilton\_vqe/10\_analytic\_direct\_representative}\\
Exhaustive check at $(14,10)$, Sec.~\ref{sec:measured} & \rf{hamilton\_vqe/14\_exhaustive\_r10\_certificate}\\
Validation, Table~\ref{tab:validation} & \rf{hamilton\_vqe/11\_analytic\_direct\_global\_validation}\\
Precision scaling, Fig.~\ref{fig:scaling} & \rf{hamilton\_vqe/12\_analytic\_direct\_precision\_scaling}\\
Noncommuting pair, Fig.~\ref{fig:stairs}d & \rf{trotter\_pair}\\
Regression suite, App.~\ref{app:compiler} & \rf{hamilton\_vqe/02\_compiler\_regression}\\
\bottomrule
\end{tabular}
\end{table}

\noindent\begin{codebox}
# Agda (Sec. 5, Sec. 7.3)
cd agda; agda -i . -i <agda-stdlib> VHamilton.agda
# Secs. 6 and 8.3
cabal run qaoa-demo -- 3 4 5
cabal run trotter-vqe-demo -- 4 0 1 2 3 4
python3 -m scripts.trotter_search.experiment pair
# Sec. 7
python -m scripts.chemistry.chem_decompose --block upper --level 14
cabal run chem-vqe -- 10 3 N exhaustive                 # N = 0, 1, 2
python -m scripts.chemistry.chem_decompose_vqe --target-error-mh T  # each T
cabal run chem-vqe -- 10 3 0 exhaustive; cabal run chem-vqe -- 10 5 0 direct
python -m scripts.experiments.chem_vqe_sweep --k-min 3 --k-max 12 \textbackslash
    --r-min 2 --r-max 6 --target-error-mh 0.1 --output-dir <folder 07>
python -m scripts.experiments.error_calculus_bounds --out-dir <folder 13>
# Sec. 8
python -m scripts.vqe.verify_chem_vqe 14 7              # step 1, exhaustive
cabal run chem-vqe -- 14 7 0 M        # M = refine, coordinate, analytic, direct
cabal run chem-vqe -- 10 5 0 direct; cabal run chem-vqe -- 14 10 0 direct
python -m scripts.experiments.validate_analytic_direct --experiment validation \textbackslash
    --out-dir <folder 11>
python -m scripts.experiments.check_coordinate_minima \textbackslash
    --matrix <folder 11>/direct_validation_matrix.csv
python -m scripts.experiments.validate_analytic_direct --experiment scaling \textbackslash
    --out-dir <folder 12>
python -m scripts.experiments.exhaustive_grid_certificate --k 14 --r 10
# App. C
cabal test regression --test-show-details=direct
\end{codebox}
\medskip

{\sloppy\noindent The folder for each command is given in Table~\ref{tab:repro}, and each folder records its own command verbatim. The figure files are generated from the same folders: Fig.~\ref{fig:errors} by \texttt{error\_calculus\_bounds}, Fig.~\ref{fig:scaling} by \texttt{scripts.experiments.direct\_scaling\_figure}, and the QAOA heatmaps of Fig.~\ref{fig:qaoa} by \texttt{scripts/figures/qaoa\_ladder\_tikz.py} from the grids in \texttt{qaoa\_one\_edge}.\par}

\end{document}